\documentclass[twocolumn]{aastex63}

\usepackage{multirow}
\usepackage{xfrac}
\usepackage{diagbox}

\newcommand{\MUSEpack}{\dataset[MUSEpack]{\doi{10.5281/zenodo.3433996}}}
\newcommand{\pyspeckit}{\texttt{pyspeckit} \citep{Ginsburg2011} }
\newcommand{\spectralcube}{\texttt{spectral\_cube}\footnote{\url{https://spectral-cube.readthedocs.io/en/latest/}} }
\newcommand{\python}{\texttt{\uppercase{python}} }

\defcitealias{GaiaCollaboration2016}{Gaia Collaboration et al. 2016}
\defcitealias{GaiaCollaboration2018}{Gaia Collaboration et al. 2018}
\defcitealias{Zeidler2018}{Paper~1}
\defcitealias{Zeidler2019a}{Paper~2}

\received{\today}
\revised{}
\accepted{}
\submitjournal{AJ}

\shorttitle{The complex internal dynamics of Wd2}
\shortauthors{Zeidler et al.}
\graphicspath{{./}{figures/}}

\begin{document}

\title{The young massive star cluster Westerlund 2 observed with MUSE. \\
	III. A cluster in motion -- the complex internal dynamics}

\correspondingauthor{Peter Zeidler}
\email{zeidler@stsci.edu}

\author[0000-0002-6091-7924]{Peter Zeidler}
\affil{Department of Physics and Astronomy, Johns Hopkins University, Baltimore, MD 21218, USA}
\affil{AURA for the European Space Agency (ESA), ESA Office, Space Telescope Science Institute, 3700 San Martin Drive, Baltimore, MD 21218, USA}

\author[0000-0003-2954-7643]{Elena Sabbi}
\affil{Space Telescope Science Institute, 3700 San Martin Drive, Baltimore, MD 21218, USA}

\author{Antonella Nota}
\affil{European Space Agency (ESA), ESA Office, Space Telescope Science Institute, 3700 San Martin Drive, Baltimore, MD 21218, USA}

\author[0000-0002-5456-523X]{Anna F. McLeod}
\affil{Centre for Extragalactic Astronomy, Department of Physics, Durham University, South Road, Durham, DH1 3LE, UK}
\affil{Department of Astronomy, University of California Berkeley, Berkeley, CA 94720, USA}
\affil{Department of Physics and Astronomy, Texas Tech University, PO Box 41051, Lubbock, TX 79409, USA}




\begin{abstract}

Analyzing the dynamical state of nearby young massive star clusters is essential understanding star cluster formation and evolution during their earliest stages. In this work we analyze the stellar and gas kinematics of the young massive star cluster Westerlund 2 (Wd2) using data from the integral field unit MUSE and complement them with proper motions from the \textit{Gaia} DR2. The mean gas radial velocity of $15.9\,{\rm km}\,{\rm s}^{-1}$ agrees with the assumption that Wd2 is the result of a cloud-cloud collision. The gas motions show the expansion of the \ion{H}{2} region, driven by the radiation from the many OB stars in the cluster center. The velocity profile of the cluster member stars reveal an increasing velocity dispersion with decreasing stellar mass and that the low-mass stars show five distinct velocity groups. Based on their spatial correlation with the cluster's two clumps, we concluded that this is the imprint of the initial cloud collapse that formed Wd2. A thorough analysis of the dynamical state of Wd2, which determines a dynamical mass range of $M_{\rm dyn,Wd2} = (7.5 \pm 1.9)\cdot10^4 - (4.4 \pm 1.1)\cdot10^5\,{\rm M}_\odot$ and exceeds the photometric mass by at least a factor of two leads to the conclusion that Wd2 is not massive enough to remain gravitationally bound. Additionally we also identify 22 runaway candidates with peculiar velocities between 30 and $546\,{\rm km}\,{\rm s}^{-1}$.

\end{abstract}



\section{Introduction}
\label{sec:introduction}

The detailed study of young star cluster (YSC) kinematics is vital to understand their long-term evolution and survivability. Only if star clusters are massive enough they may overcome the so-called ``infant mortality'' \citep{Lada2003} surviving internal and external processes that disturb the gravitational potential, such as supernova explosions, which lead to abrupt, violent gas expulsion \citep[e.g.,][]{Goodwin2006,Bastian2006,PortegiesZwart2010}, the interaction or collision with a giant molecular cloud (GMC) in the Galactic disk or a change in the Galactic tidal field \citep[e.g.,][and references therein]{Krumholz2020}. The detailed kinematic analysis of such young, and still embedded clusters and their gaseous and stellar content down to the hydrogen burning limit (or even below) is challenging and has only become feasible in recent years with new telescopes, instruments, and computational methods. For example, \citet{Zari2019a} showed the three dimensional, highly substructured nature of the Orion Nebular Cloud (ONC) detecting multiple kinematic components with distinct age differences, while \citet{Jerabkova2019} suggest the detection of multiple populations in its young stellar population. \citet{McLeod2015} studied the detailed kinematics of the ``Pillars of Creation'' in the Eagle Nebula finding radial velocity (RV) differences of $\sim 2\,{\rm km}/{\rm s}$ between the different pillars. Their analysis also revealed a possible protostellar outflow and let them identify both lobes as a blue and a red shifted counterpart. Time domain studies \citep[e.g.,][]{Sabbi2020} have the capability of detecting protoplanetary disks, stellar variability, and the binary fraction, which is important to the evolution of the whole cluster system  (we refer to \citet{PortegiesZwart2010} and \citet{Krumholz2020} for a detailed overview).

The measurement of the internal kinematics of YSCs has been observationally very expensive. Especially for the determination of RVs, high resolution stellar spectra had to be obtained. For many fiber or slit spectrographs, only a handful of stars can be observed simultaneously. With the development of efficient, large field of view (FOV) integral field units (IFUs) over the past decade and the increasing computational capacities it has become possible to study the RV of entire stellar populations in resolved star clusters. The optical IFU with the largest FOV to date ($1\,{\rm arcmin}^2$) is the Multi Unit Spectroscopic Explorer \citep[MUSE,][]{Bacon2010} mounted at UT4 of the Very Large Telescope (VLT), which allows us to survey larger regions similar to photometric studies. MUSE has been proven to be an excellent instrument to spectroscopically map nearby star-forming regions to study the kinematics of their stars and the gas simultaneously \citep[e.g.,][for the latter two, hereafter Paper~1 and Paper~2]{McLeod2015,McLeod2020,Zeidler2018,Zeidler2019a}. In \citetalias{Zeidler2018} we showed that it is indeed possible to measure stellar RVs in YSCs to an accuracy of $\sim 2\,{\rm km}\,{\rm s}^{-1}$ using \MUSEpack, despite the lack of pre-main sequence (PMS) stellar spectral libraries and the variable and high local background. For a detailed description of the code and the assessment of the uncertainties we refer to \citetalias{Zeidler2019a}.

This work is the third paper in a series \citepalias{Zeidler2018,Zeidler2019a} spectroscopically studying the Galactic young massive star cluster Westerlund 2 \citep[Wd2,][]{Westerlund1961} using MUSE data. Wd2 is the central ionizing star cluster of the \ion{H}{2} region RCW49 \citep{Rodgers1960} located in the Carina-Sagittarius spiral arm at a distance of $\sim4.16$\,kpc \citep{Zeidler2015,VargasAlvarez2013} at an age of 1--2\,Myr \citep{Zeidler2015}. With a total photometric stellar mass of $3.7\cdot10^4\,{\rm M}_\odot$, Wd2 is the second most massive young star cluster in the Milky Way \citep[MW,][]{Zeidler2017}, after Westerlund~1 \citep[$\sim5\cdot10^4\,{\rm M}_\odot$, e.g.,][]{Clark2005,Andersen2017}. This cluster is built from two coeval clumps, the MC and NC \citep{Hur2014,Zeidler2015} and is highly mass segregated \citep{Zeidler2017}. The close proximity, its numerous OB stars in the cluster center \citep[e.g.,][]{Rauw2004,Rauw2011,Bonanos2004,VargasAlvarez2013}, and its young age (so far no supernova explosion has been detected) make Wd2 a prime target to study the internal processes of star cluster formation and pre-supernovae evolution.

In this work we present a detailed analysis of the dynamical state of Wd2 to determine whether it has a chance to overcome infant mortality, and to better understand its formation process. This paper is structured the following. In Sect.~\ref{sec:data} we give a brief introduction to the used data. In Sect.~\ref{sec:spat_dist} we reanalyze the spatial structure of Wd2 to obtain missing key parameters. In Sect.~\ref{sec:RV} we present a detailed analysis of the gas and stellar RVs including the dynamical state of Wd2. Sect.~\ref{sec:Gaia} introduces the data obtained from the \textit{Gaia} mission and in Sect.~\ref{sec:runaways} we analyze high-velocity stellar runaway candidates. In Sect.~\ref{sec:discussion} we provide a in-depth discussion about the results obtained in the previous sections and put them into a greater context, while in Sect.~\ref{sec:summary} we summarize the analysis and our findings.
 
\section{The data and data reduction}
\label{sec:data}

We will only provide a brief overview of the dataset, data reduction, and RV measurements. A detailed description was presented in \citetalias{Zeidler2018} and \citetalias{Zeidler2019a}. The data and their derived products used in this work, such as stellar RVs, are identical to those of \citetalias{Zeidler2019a}.

We surveyed Wd2 using 21.5\,h of VLT/MUSE time (Program ID: 097.C-0044(A), 099.C-0248(A), PI: P.~Zeidler). We combined 11 short (220\,s) and 5 long (3600\,s) exposures to simultaneously cover the gas, the high-mass, luminous OB stars, and the fainter PMS stars down to $\sim 1\,{\rm M}_\odot$. MUSE was operated in the extended mode covering a wavelength range of 4600--$9350\,{\rm \AA}$. The short exposures were executed in the wide-field mode without the adaptive optics (AO) system (WFM\_NOAO) while four of the five long exposures were executed in the wide-field mode with AO (WFM\_AO), which results in a spatial resolution improvement by a factor of two. Because in AO mode the notch filter of the Na-lasers blocks the coverage in the 5780--$5990\,{\rm \AA}$ range, we chose the WFM\_NOAO mode for the short exposures as to cover the \ion{He}{1}$\lambda$5876 line.

The data was reduced with the \texttt{musereduce} module of the python package \MUSEpack\footnote{\dataset[MUSEpack]{\doi{10.5281/zenodo.3433996}} is made available for download on Github \url{https://github.com/pzeidler89/MUSEpack.git}} \citep{Zeidler2019} together with the standard MUSE data reduction pipeline \citep{Weilbacher2012,Weilbacher2015}. In total we extracted 1726 stellar spectra with a mean signal-to-noise ratio\footnote{Whenever we refer to the S/N of spectra we always provide a S/N per spectral bin.} (S/N) $\ge5$ using the software package PampelMuse \citep{Kamann2013,Kamann2016} in combination with our deep, high-resolution, multi-band photometric star catalog extracted from \textit{Hubble} Space Telescope (HST) observations \citep[ID: 13038, PI: A. Nota,][]{Zeidler2015} to detect and de-blend the stellar spectra. The world coordinate system (WCS) of all data were corrected to match the \textit{Gaia} data release 2 \citepalias[DR2,][]{GaiaCollaboration2016,GaiaCollaboration2018}.

\section{The stellar spatial distribution}
\label{sec:spat_dist}
\citet{Zeidler2015} confirmed the finding of \citet{Hur2014} that Wd2 is built from two subclumps, the main cluster (MC) and the northern clump (NC) with a projected separation of $\sim 1$\,pc at the distance of Wd2. We reanalyze the spatial structure using both the stellar surface density and the stellar mass density obtained from our photometric HST catalog down to the 50\% completeness limit of all cluster member stars \citep{Zeidler2017}. We use a maximum likelihood approach to fit two peaks to the density distributions including a common offset to account for a halo of lower-mass stars and test two different distributions: 1) two 2D Gaussian profiles and 2) the Elson-Fall-Freeman (EFF) profile \citep{Elson1987}. The latter is an empirical surface density profile as a function of $r$ that was found to well describe the surface density of massive YSCs in the MW. It has the form:

\begin{equation}
\label{eq:EFF}
    \Sigma(r) = \Sigma_0 \left(1+\frac{r^2}{a^2}\right)^{-\sfrac{\gamma}{2}},
\end{equation}

with $\Sigma_0$ being the peak surface density and $a$ being a scale parameter. The core radius, $r_c$, used by the \citet{King1966} profile (to fit Globular Cluster profiles) is:

\begin{equation}
\label{eq:EFF_rc}
    r_c = a \left(2^{\sfrac{2}{\gamma}} -1\right)^{\sfrac{1}{2}},
\end{equation}

where $\gamma$ and $a$ are the EFF profile parameters \citep[for a detailed summary see also][]{PortegiesZwart2010}.

\begin{figure*}[htb]
\includegraphics[width=0.95\textwidth]{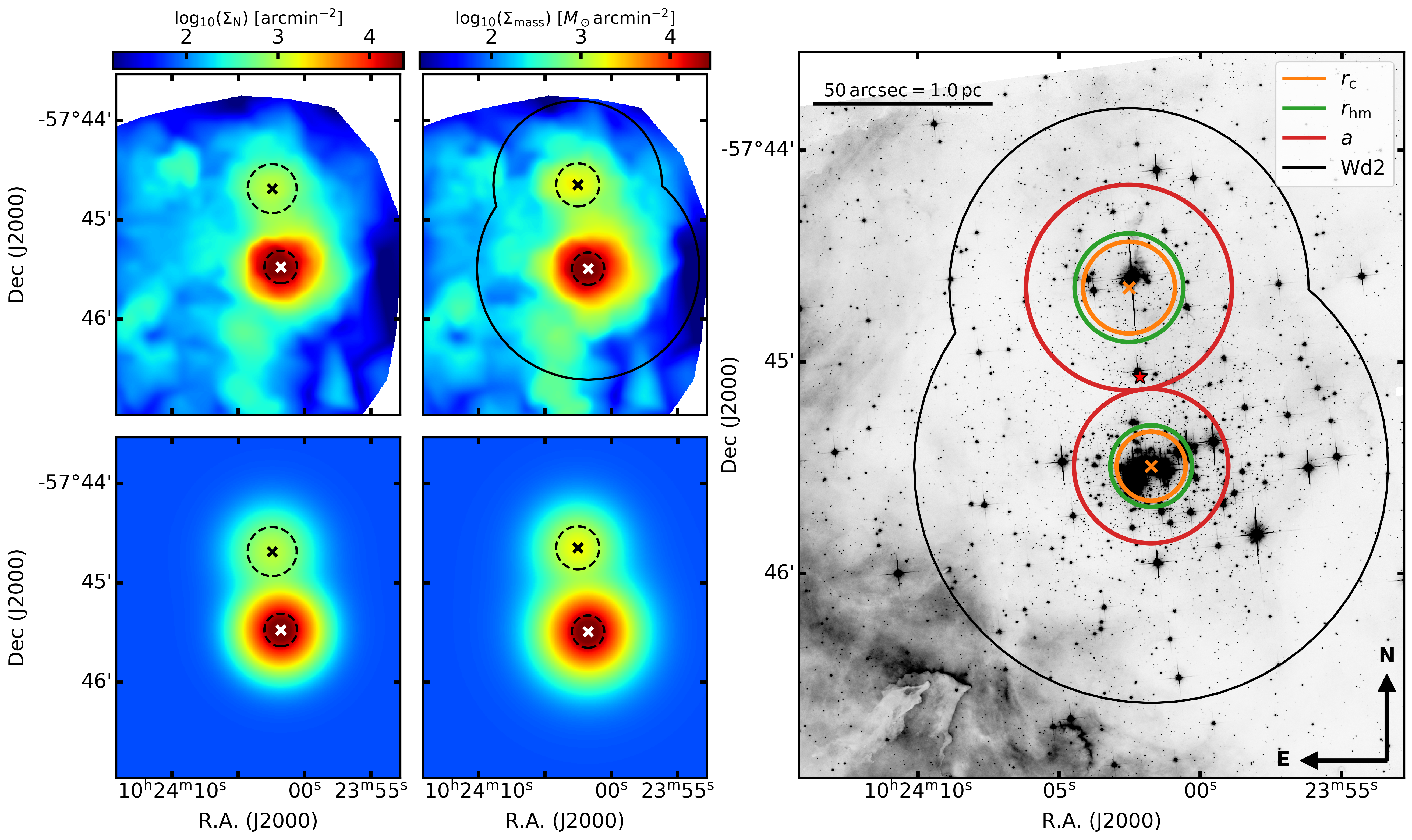}
\caption{The star and mass density distributions of the completeness corrected photometric star catalog of Wd2 down to the 50\% completeness limit. On the top left are the observed stellar mass and number density distributions and on bottom left are shown the simulated density distributions. The core radii for the MC and NC are indicated by the dashed circles. On right we show the HST $F814W$ image with the core radii $r_{\rm c}$, the half-mass radii $r_{\rm hm}$, and the scale parameters $a$, as well as the MC and NC centers of the best-fit EFF model for the mass density. The red asterisk marks the center of Wd2 defined by the geometric mean between the MC and the NC. The black outline marks the cluster region of Wd2.}
\label{fig:spat_dist}
\end{figure*}

After running extensive Markov-Chain Monte Carlo (MCMC) fitting of the two density distributions to the data, the Akaike information criterion \citep[AIC, ][]{Akaike1974}, the Bayesian information criterion \citep[BIC, ][]{Schwarz1978}, and the Watanabe -- Akaike information criterion \citep[WAIC, ][]{Watanabe2010,Gelman2013} clearly favors the EFF model over a Gaussian distribution. The best-fit parameters for the mass and number density distributions are shown in Tab.~\ref{tab:spat_dist} and Fig.~\ref{fig:spat_dist}.

\begin{deluxetable*}{ccccllrrr}[htb]
	\tablecaption{The best-fit surface density parameters \label{tab:spat_dist}}
	\tablehead{\multicolumn{1}{c}{ } & \multicolumn{1}{c}{Region} & \multicolumn{1}{c}{R.A.} & \multicolumn{1}{c}{Dec.} & \multicolumn{1}{c}{$\Sigma_0$} & \multicolumn{1}{c}{$\Sigma_{bck}$} & \multicolumn{1}{c}{$a$} & \multicolumn{1}{c}{$r_c$} & \multicolumn{1}{c}{$\gamma$} \\
	\multicolumn{2}{c}{ } &\multicolumn{1}{c}{(J2000)} &\multicolumn{1}{c}{(J2000)} & \multicolumn{1}{c}{$({\rm arcmin}^{-2})$}  & \multicolumn{1}{c}{$({\rm arcmin}^{-2})$}  & \multicolumn{1}{c}{(pc)} & \multicolumn{1}{c}{(pc)} & \multicolumn{1}{c}{}
		}
	\startdata
	\multirow{2}{*}{$\Sigma_{\rm num}$} & MC & $10^{\rm h}24^{\rm m}01^{\rm s}.788$   & $-57^{\circ}45^{\rm m}28^{\rm s}.63$ & $3.02 \cdot 10^4$  & \multirow{2}{*}{68.27} & $0.53 \pm 0.01$ & $0.20 \pm 0.01$ & $10.47 \pm 0.01$\\
	 & NC & $10^{\rm h}24^{\rm m}02^{\rm s}.438$ & $-57^{\circ}44^{\rm m}41^{\rm s}.28$  & $1.00 \cdot 10^3$  & & $0.78  \pm 0.01$ &  $0.30  \pm 0.02$ &  $10.07  \pm 0.05$ \\[0.2cm]
	\multirow{2}{*}{$\Sigma_{\rm mass}$} & MC & $10^{\rm h}24^{\rm m}01^{\rm s}.735$ & $-57^{\circ}45^{\rm m}29^{\rm s}.60$ & $3.72 \cdot 10^4\,{\rm M}_\odot$  & \multirow{2}{*}{$70.82\,{\rm M}_\odot$} & $0.44 \pm 0.01$ &$0.20 \pm 0.01$ & $7.61 \pm 0.01$\\
	 & NC &  $10^{\rm h}24^{\rm m}02^{\rm s}.521$ & $-57^{\circ}44^{\rm m}39^{\rm s}.00$ & $1.86 \cdot 10^3\,{\rm M}_\odot$  & & $0.59 \pm 0.01$  & $0.26 \pm 0.01$ &  $7.63 \pm 0.02$\\
	\enddata
	\tablecomments{The sub clump parameters of the best fit EFF model based on the number density (first two rows) and the mass density (second two rows), for the MC and NC. At a distance of 4.16\,kpc, a projected distance of 50\,arcsec are 1\,pc.}
\end{deluxetable*}

The dynamical evolution of a star cluster is highly driven by its mass distribution and, therefore, we will use the mass density as reference distribution. We define the coordinates of the Wd2 cluster (${\rm R.A.} = 10^{\rm h}24^{\rm m}02^{\rm s}.128$, ${\rm Dec.} = -57^{\circ}45^{\rm m}04^{\rm s}.30$) as the geometric mean of the centers of the two clumps, similar to the definition in \citet{Zeidler2015}. Integrating over the mass density distribution leads to a total clump mass above the 50\% completeness limit of $m_{\rm MC}^{50} = (0.55 \pm 0.01)\cdot 10^4\,{\rm M}_\odot$ and $m_{\rm NC}^{50} = (0.05 \pm 0.01)\cdot 10^4 \,{\rm M}_\odot$, which agrees with the masses estimated using the stellar mass function \citep{Zeidler2017}.

The half-mass radius for the EFF profile is defined as:

\begin{equation}
\label{eq:EFF_hm}
    r_{\rm hm} = a \left(0.5^{\frac{2}{2-\gamma}} -1\right)^{\sfrac{1}{2}},
\end{equation}

and yields $r_{\rm hm} = (0.23 \pm 0.01)\,{\rm pc}$ and $r_{\rm hm} = (0.31 \pm 0.01)\,{\rm pc}$ for the MC and NC, respectively\footnote{For the detailed error propagation see eq.~\ref{eq:sEFF_mass_tot} to \ref{eq:sEFF_rhm}}. The exponential decline of the stellar distribution is with $\gamma_{\rm MC} = 7.61 \pm 0.01$ and $\gamma_{\rm NC} = 7.63 \pm 0.02 $ steeper than observed in other YSCs \citep[typical values are $\gamma=2$--3, e.g.,][]{Elson1987,PortegiesZwart2010}, which may be explained by the composite nature of Wd2, the high degree of mass-segregation, and that the distribution is only fit to the stars above the 50\% completeness limit, which increases core densities.

For any further analysis of the cluster stellar population we define the size of Wd2 as the combined, encircled area of 1.5 times the radius (around each clump)\footnote{The factor of 1.5 is chosen such that the irregularly, elongated shape of the stellar distribution (see top, left frames of Fig.~\ref{fig:spat_dist}) is taken into account.}, at which the stellar mass density drops to the halo (background) density of $\Sigma_{\rm bck} = 70.82\,{\rm M}_\odot$ (black lines in Fig.~\ref{fig:spat_dist}).

\section{The radial velocity profile}
\label{sec:RV}
To measure RVs\footnote{Throughout the paper we may use the term ``velocity'' interchangeable for ``radial velocity'' if it is clear from context.} we used our new method that allows us to measure stellar RVs without the need of a spectral template library, which was implemented in the \texttt{RV\_spectrum} module of \MUSEpack. \texttt{RV\_spectrum} uses strong stellar absorption lines in combination with a Monte Carlo approach to measure stellar RVs to an accuracy of $1.10\,{\rm km}\,{\rm s}^{-1}$. A detailed description of \MUSEpack, the measurements of the stellar RVs and the underlying assumptions and sources of RV uncertainties are provided in \citetalias{Zeidler2019a}.

\subsection{\texttt{RV\_spectrum} -- a new way to measure RVs}
\label{sec:RV_spectrum}

To aid the reader in understanding the further analyses, we provide a brief summary of the key steps for measuring RVs with the \texttt{RV\_spectrum} class of\MUSEpack:

\begin{enumerate}
    \item To provide a clean sample of spectra, a visual inspection of all extracted spectra is necessary. It ensures that the local background subtraction was successful and that the spectral lines used for the fit do not show signs of emission, which is common for PMS stars and is the result of accretion processes.
    
    \item The regions around each of the chosen absorption lines are fitted, using a user-provided spectral line library, together with a low-order polynomial to match the local continuum. A spectral template is created using the line parameters of the best-fitting solution and the rest-frame wavelengths.
    
    \item These templates are cross-correlated with the stellar spectra using the core of each line, which provides a RV measurement per absorption line. This cross-correlation is typically repeated 10,000 times and for each iteration the uncertainties of the spectrum are randomly reordered. A sigma clipping is applied that ensures that lines with ``odd'' profiles are removed from the final RV fit.
    
    \item The remaining, trustworthy, lines are now cross-correlated together with a typical repetition of 20,000 times. The resulting Gaussian distribution gives the RV of the star (mean) and the uncertainty ($1\sigma$).

\end{enumerate}

Extensive tests of this method are described in \citetalias{Zeidler2019a} to ensure its reliability and to show its limitations and possible sources for errors. In Appendix~\ref{sec:plots} we show the extracted and fitted spectra of four different Wd2 member stars (see Fig.~\ref{fig:spec_fit_HeIHeII_9183}, \ref{fig:spec_fit_HeIHeII_7613}, and \ref{fig:spec_fit_MgICaII}.)

\subsection{The gas velocities}
\label{sec:RV_gas}

\begin{figure*}[htb]
\includegraphics[width=\textwidth]{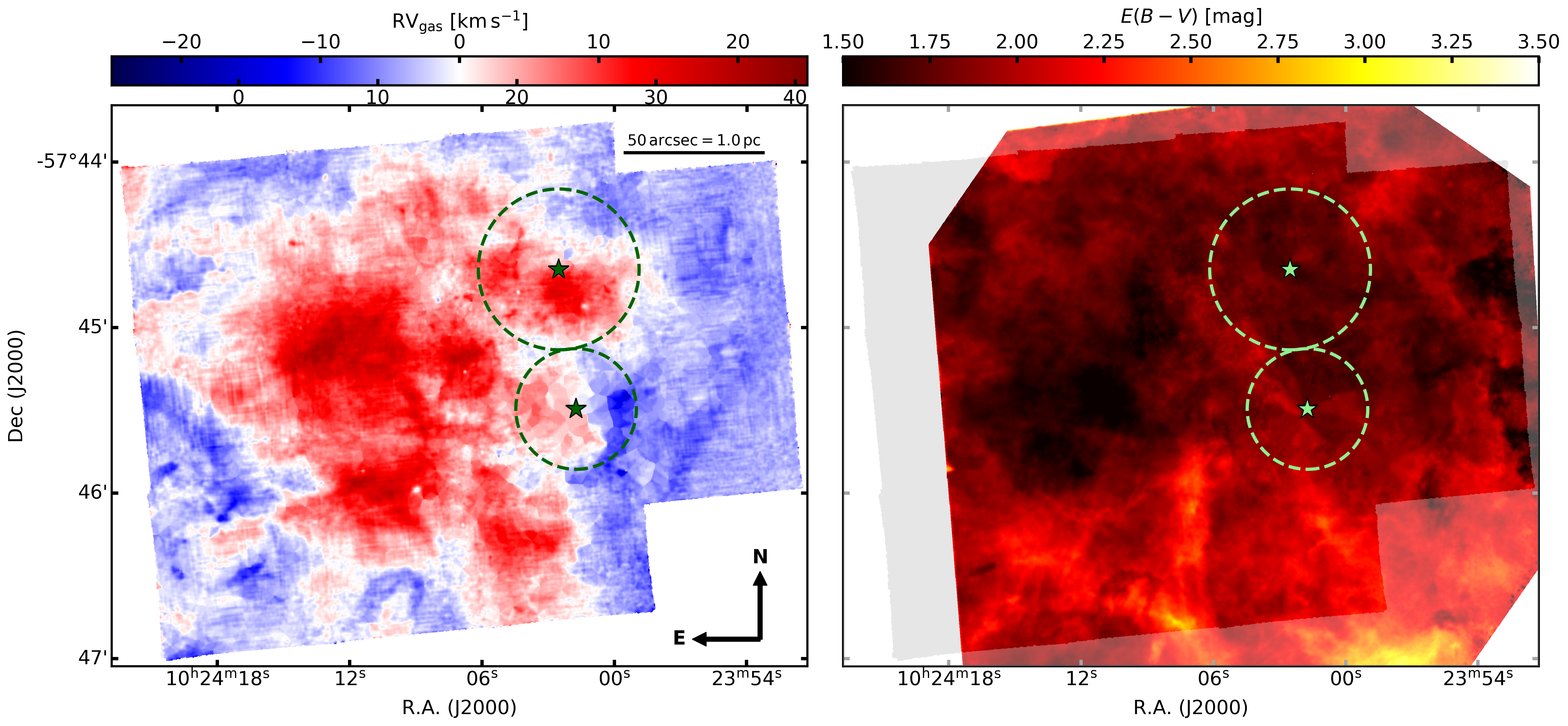}
\caption{Left: The RV map of the gas. The bottom numbers of the color bar represent the measured gas RVs while the top numbers mark the RVs corrected for the systemic motion of Wd2 ($15.9\,{\rm km\,s}^{-1}$). Right: The $E(B-V)$ color excess map \citep[similar to][]{Zeidler2015} at a resolution of 0.8\,arcsec representing the average seeing of the MUSE dataset (Paper 2). The outline of the gas RV map is over plotted to orient the reader. In both frames are marked the centers of the MC and the NC including their scale parameters $a$ in green as defined in Sect.~\ref{sec:spat_dist}.}
\label{fig:EBV_RVgas}
\end{figure*}

To obtain the gas velocity profile we use a similar approach as \citet{McLeod2015} by stacking the spectra of individual gas emission lines to a single spectral line per spatial pixel (spaxel). This stacking results in a well-sampled line, which is fit by a Gaussian profile to measure the RVs on a spaxel-by-spaxel basis. This method is computationally more efficient than measuring RVs with \MUSEpack~ but it is only applicable if there are a significant number of strong, non-blended spectral lines available that have a reasonably flat continuum. We used the \python packages \pyspeckit and \spectralcube to combine the H$
\alpha$, the \ion{N}{2}\,$\lambda\lambda6549.85, 6585.28\,{\rm \AA}$, and the \ion{S}{2}\,$\lambda\lambda6718.29, 6732.67\,{\rm \AA}$ emission lines. The continuum was extracted in the spectral range of 6620--$6660\,{\rm \AA}$. We avoid using the [\ion{O}{1}]\,$\lambda\lambda 6300, 6363{\rm \AA}$ emission lines due to the applied ``modified sky subtraction'' \citepalias{Zeidler2019a}, which properly recovers these lines but it may slightly change their centroids due to Telluric residuals. To obtain the mean gas velocity of the \ion{H}{2} region small differences in the velocities of the individual gas components can be neglected (e.g., \citealp{McLeod2016}, \citetalias{Zeidler2018}). Additionally, we masked the stars and extrapolated the gas velocities at each stellar position to get a clean gas velocity map (see left frame of Fig.~\ref{fig:EBV_RVgas}). The median RV of the gas, determined from all MUSE pixels, is $15.9\,{\rm km\,s}^{-1}$, which we will use henceforth as the systemic RV of Wd2 relative to the Sun. This systemic velocity is subtracted from all further RV measurements in this study unless stated otherwise.

\begin{figure*}[htb]
\includegraphics[width=\textwidth]{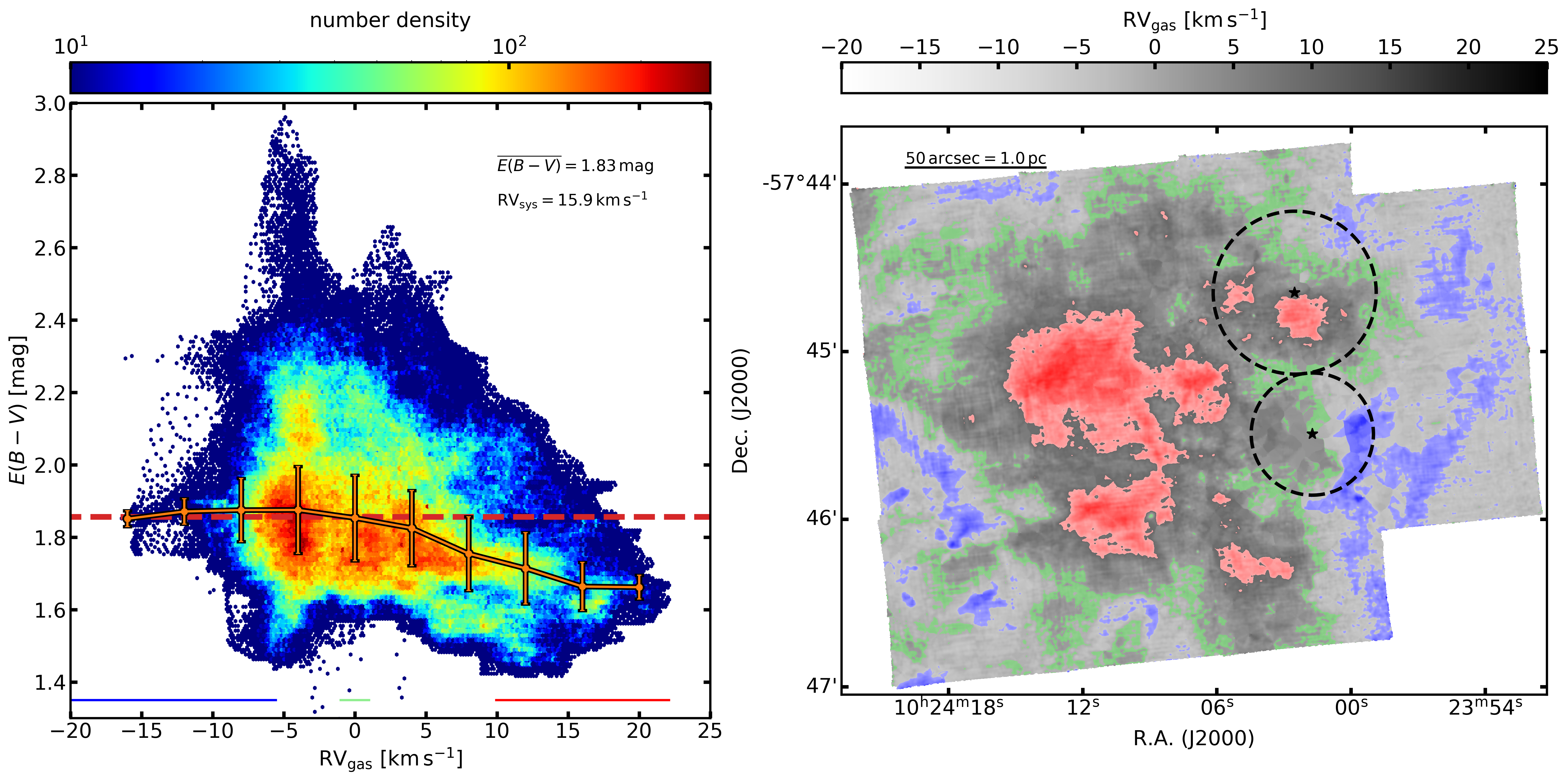}
\caption{Left: The pixel-by-pixel $E(B-V)$ color excess vs. the gas RVs. In orange we mark the velocity binned extinction average showing a clear correlation between the two. The red-dashed line shows the median extinction. Right: The gas RV map marked with the bottom (blue) and the top (red) 10\% of the RV distribution as well as, the gas with $v_{\rm sys}\pm 1\,{\rm km}\,{\rm s}^{-1}$. These RV ranges are also marked on the bottom of the left frame. The center of the clumps as well as the scale radii $a$ (dashed circles) are plotted to orient the reader.}
\label{fig:EBV_RVgas_analysis}
\end{figure*}

The gas velocity profile (left frame of Fig.~\ref{fig:EBV_RVgas}) clearly shows that the central part of the cloud is receding while the outer ridges move toward us. When comparing the gas RV map with the extinction map \citep[see right frame of Fig.~\ref{fig:EBV_RVgas} and][]{Zeidler2015}, we see a correlation between the magnitude of the $E(B-V)$ color excess and the gas motion. By comparing the average gas velocity with the average $E(B-V)$ color excess (left frame of Fig.~\ref{fig:EBV_RVgas_analysis}) we indeed see significantly higher $E(B-V)$ values at negative RVs. Locations with a lower line-of-sight extinction allow us to look deeper into the gas cloud\footnote{We note here that the \citet{Zeidler2015} color-excess map does not distinguish between the extinction caused by the \ion{H}{2} region and the foreground extinction. Given the relatively small FOV ($\sim 5' \times 5'$) of the survey area, we do not expect any significant variations of the foreground extinction. The regions with low extinction in the \citet{Zeidler2015} color-excess map are in agreement with the foreground color-excess, $E(B-V)_{\rm fg}=1.05$\,mag, estimated by \citet{Hur2014}.}. We conclude that we actually see the expansion of the \ion{H}{2} region driven by the stellar winds and the far ultraviolet (FUV) radiation of the numerous OB stars of Wd2 (e.g., \citealp{Rauw2004,Rauw2005,Rauw2007,Bonanos2004,VargasAlvarez2013,Drew2014}, \citetalias{Zeidler2018}). To better visualize the expansion we show a black-white version of the gas RV map (right frame of Fig.~\ref{fig:EBV_RVgas}) in which we marked the bottom (blue) and the top (red) 10\% of the RV distribution, as well as, the gas with $v_{\rm sys}\pm 1\,{\rm km}\,{\rm s}^{-1}$. This suggests a differential RV of the \ion{H}{2} region of $\sim 15\,{\rm km\,s}^{-1}$.

\subsection{The stellar radial velocities}
\label{sec:RV_stars}
To measure stellar RVs we used the following spectral absorption lines, depending on the stellar type: \ion{He}{1}\,$\lambda\lambda\,4922, 5876, 6678, 7065\,{\rm \AA}$, \ion{He}{2}\,$\lambda\lambda\,4685, 5412\,{\rm \AA}$, \ion{Mg}{1}\,$\lambda\lambda\,5367, 5172, 5183\,{\rm \AA}$, \ion{Na}{1}\,$\lambda\lambda\,5889, 5895\,{\rm \AA}$, and \ion{Ca}{2}\,$\lambda\lambda\,8498, 8542, 8662\,{\rm \AA}$. We intentionally avoided other strong absorption lines, such as Balmer lines since these may be unreliable for RV measurements due to the young stellar age, the possible ongoing accretion processes, and nebular contamination. An overview of the applied method is given in Sect.~\ref{sec:RV_spectrum} and an in-depth analysis of the underlying assumptions, the selection criteria, as well as the limitations and uncertainties is presented in \citetalias{Zeidler2019a}.

\begin{figure}[htb]
\plotone {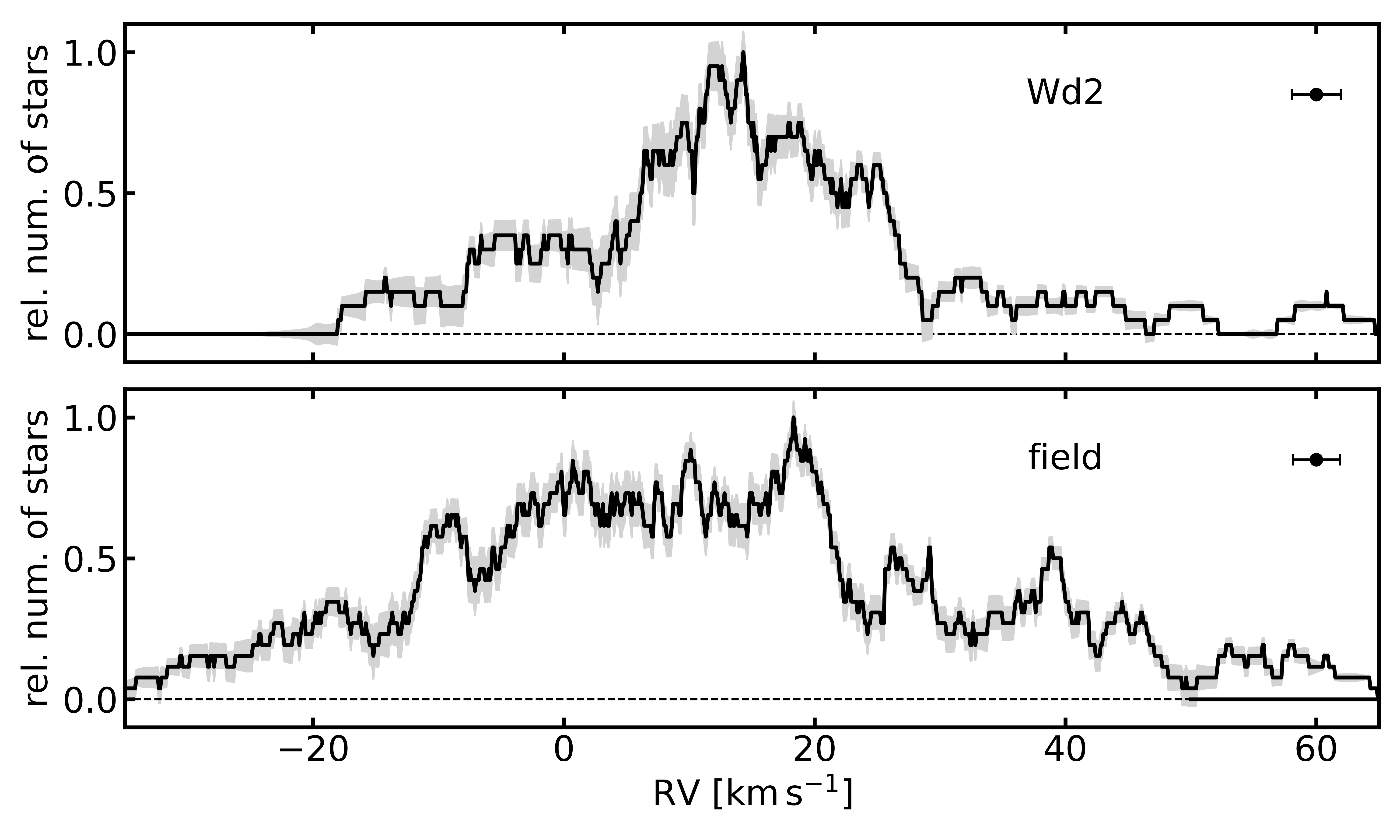}
\caption{The RV distribution of the cluster members (top panel) and field stars (bottom panel). The step size is $0.1\,{\rm km}\,{\rm s}^{-1}$ with a bin width using the typical uncertainties of $\sigma_{\rm typ}^{\rm clm} = 1.96\,{\rm km}\,{\rm s}^{-1}$ and $\sigma_{\rm typ}^{\rm field} = 1.87\,{\rm km}\,{\rm s}^{-1}$.}
\label{fig:RV_dist}
\end{figure}

In total we extract reliable RVs from 388 stars. Based on the $F814W-F160W$ vs. $F814W$ and the $F555W-F814W$ vs. $F555W$ color-magnitude diagrams (CMDs) created with our HST photometric catalog \citep{Zeidler2015}, 117 source are located in the Wd2 cluster and 271 are foreground field stars. The high extinction ($A_V=6.12$\,mag) toward Wd2 allows for a clean separation between cluster members and field stars. The typical (mean) RV uncertainties are $\sigma_{\rm typ}^{\rm clm} = 1.96\,{\rm km}\,{\rm s}^{-1}$ and $\sigma_{\rm typ}^{\rm field} = 1.87\,{\rm km}\,{\rm s}^{-1}$ for the cluster members and MW field stars, respectively. In Fig.~\ref{fig:RV_dist} we show the stellar RV distribution of the cluster members (top panel) and field stars (bottom panel). The field stars span over a wider RV range than the cluster members but generally their RV spaces overlap. This is expected due to the location of Wd2 close to the tangent point of the Carina-Sagittarius spiral arm. To create the RV histograms we use a running mean with a step size of $0.1\,{\rm km}\,{\rm s}^{-1}$ and a bin width of the typical uncertainty. This method reduces a possible bias caused by binning the data.

\subsection{The Wd2 velocity profile}
From the isochrone fitting to CMDs \citep{Zeidler2015,Sabbi2020} we know that the PMS turn-on is at $\sim 3-5\,{\rm M}_\odot$, which means that most O and B stars are already in their main-sequence (MS) phase. Therefore, we divide the stars into three groups:

\begin{itemize}
    \item[1.] O-stars: showing \ion{He}{1} and \ion{He}{2} absorption features (16 stars),
    \item[2.] B-stars: showing \ion{He}{1} but no \ion{He}{2} absorption features (26 stars), and
    \item[3.] later type stars: showing metal features, such as \ion{Mg}{1}-Triplet or \ion{Ca}{2}-Triplet (75 stars).
\end{itemize}

\begin{figure*}[htb]
\plotone {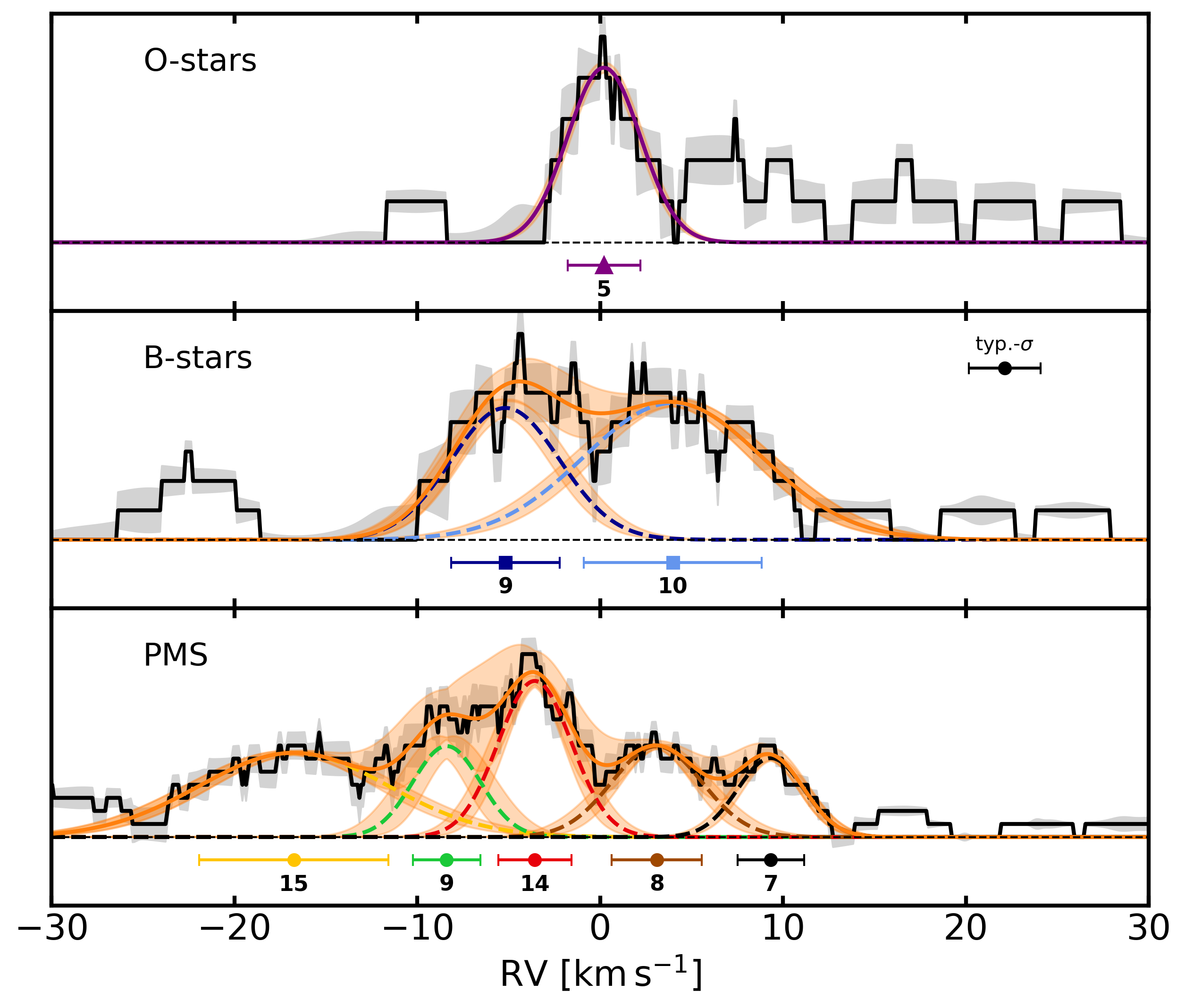}
\caption{The normalized RV distribution of the cluster member O-stars (top panel), B-stars (middle panel), and later-type PMS stars (bottom panel). The histograms are created in the same way as Fig.~\ref{fig:RV_dist}. The orange line represents the cumulative RV distribution. The results of the MCMC fits are shown by the dashed Gaussians. At the bottom of each panel we mark the mean RV as well as the velocity dispersion of each RV group. The numbers indicate the bona-fide stars per RV group (${\rm rv}\pm1\sigma$).}
\label{fig:RV_dist_grouped}
\end{figure*}

These groups divide the stars into different evolutionary stages and certain mass ranges. As the next step we create RV histograms of the three groups (see Fig.~\ref{fig:RV_dist_grouped}) with the same method as Fig.~\ref{fig:RV_dist}. 

The histograms immediately reveal a significant difference in the velocity distribution of the stellar types. The more massive the stars, the smaller their velocity dispersion. To quantify this initial analysis we fit a combination of Gaussians to the RV histograms using MCMC. This allows us to properly account for the individual RV uncertainties. The three distributions are best described by a single Gaussian for the O-stars, a combination of two Gaussians for the B-stars, and five Gaussians for the PMS stars. Using the AIC and BIC as well as the convergence of the MCMC fit we ensure that five velocity groups are the best fitting number of components without over fitting the distribution. For completeness, the corresponding corner plots can be found as Fig.~\ref{fig:corner_obstars} and \ref{fig:corner_pms} in the Appendix~\ref{sec:plots}. The results of the fits are also shown in Fig.~\ref{fig:RV_dist_grouped} and listed in Tab.~\ref{tab:RV_stars_comp}. The uncertainties are represented by one standard deviation of the marginalized distributions reflecting the contribution of the individual stellar RV uncertainty measurements. The inspection of the O-star histogram shows a possible second peak at $\sim (9\pm4)\,{\rm km}\,{\rm s}^{-1}$ but the fit does not converge. 

\begin{deluxetable}{lrrrr}[htb]
	\tablecaption{The stellar RV components \label{tab:RV_stars_comp}}
	\tablehead{\multicolumn{1}{c}{name} & \multicolumn{1}{c}{RV} & \multicolumn{1}{c}{$\sigma$RV} & \multicolumn{2}{c}{n (stars)} \\
	\multicolumn{1}{c}{} & \multicolumn{1}{c}{(${\rm km}\,{s}^{-1}$)} &\multicolumn{1}{c}{(${\rm km}\,{s}^{-1}$)} & \multicolumn{1}{c}{all} & \multicolumn{1}{c}{Wd2}}
	\startdata
	\multicolumn{5}{c}{O-stars} \\
     O1 (purple) & $0.21\pm0.08$ & $1.99\pm0.08$ & 5 & 5\\[0.1cm]
     \multicolumn{5}{c}{B-stars} \\
     B1 (dark blue) & $-5.18\pm0.24$& $2.96\pm0.21$ & 9 & 9\\
     B2 (light blue) & $3.90\pm0.36$  & $4.86\pm0.35$ &  10 & 9\\[0.1cm]
     \multicolumn{5}{c}{PMS-stars} \\
     PMS1 (yellow) & $-16.75\pm0.33$  & $5.18\pm0.35$ &  15 & 12\\
     PMS2 (green) & $-8.40\pm0.38$  & $1.85\pm0.33$ &  9 & 6\\
     PMS3 (red) & $-3.57\pm0.19$  & $2.00\pm0.22$ &  14 & 14\\
     PMS4 (brown) & $3.10\pm0.16$  & $2.46\pm0.36$ &  8 & 5\\
     PMS5 (black) & $9.33\pm0.14$  & $1.81\pm0.14$ &  7 & 5\\
	\enddata
	\tablecomments{The results of the MCMC fit to the stellar RV distributions. Column 1 is the velocity group name used throughout the rest of this work. The indicated colors correspond to the ones used in the figures. Column 2 shows the mean RVs while Column 3 shows the velocity dispersion of each component. Column 4 and 5 are the number of stars located within $1\sigma$ of each RV component inside the survey area and inside the Wd2 cluster (as defined in Sect.~\ref{sec:spat_dist}).}
\end{deluxetable}

To investigate the origin of the individual stellar velocity groups we analyze the spatial location of the stars within one standard deviation of the mean of each velocity group (error bars in Fig.~\ref{fig:RV_dist_grouped}). Additionally, the stars also need to be located within the Wd2 cluster as defined in Sect.~\ref{sec:spat_dist}. These two limitations ensure that we only analyze stars that can be uniquely identified with one velocity group and whose locus coincides with the immediate cluster, which leaves us with 5 to 14 stars for each group (see Tab.~\ref{tab:RV_stars_comp}). We use a 2D kernel density estimator (KDE) with a Gaussian kernel to better visualize the number density of the (relatively low number of) stars in each velocity group (see Fig.~\ref{fig:kde}).

\subsubsection{The O and B stars}
The O and B stars are mostly concentrated toward the center of Wd2 with the majority co-located with the MC. This is in complete agreement with a highly mass-segregated cluster. There is no apparent correlation between the spatial location of the stars and the two velocity groups of the B stars. Possible undetected binaries can be excluded as a source for the two peaks of the B-star distribution. The dispersion of MUSE is 2.4\,\AA, which means that a minimum relative velocity of $\sim 80$--$160\,{\rm km}\,{\rm s}^{-1}$ is necessary to detect line splitting caused by the binary components (compared to the $(9.08\pm0.43)\,{\rm km}\,{\rm s}^{-1}$ between B1 and B2, see Tab.~\ref{tab:RV_stars_comp}). The visual inspection of all used spectra does not hint for any such line splitting.

\subsubsection{The late-type PMS stars}
\begin{figure*}[htb]
\plotone{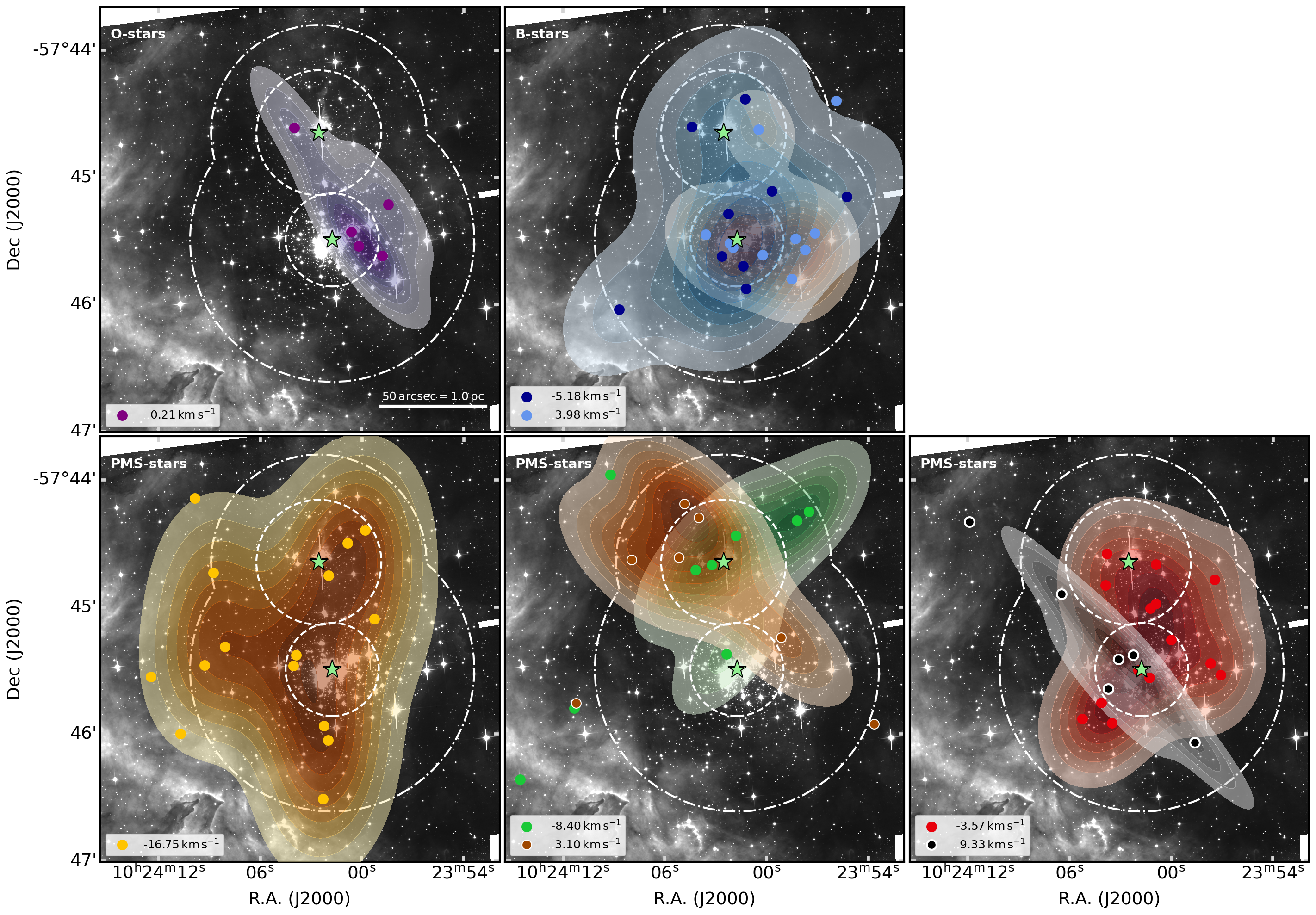}
\caption{The spatial distribution of the individual RV groups for the O and B stars (top frames) and the later-type PMS stars (bottom three frames). The colors scheme is identical to Fig.~\ref{fig:RV_dist_grouped} for the velocity groups. We divided the later-type PMS stars into three individual plots to avoid confusion. The contours represent the spatial stellar density of each stellar velocity group determined via a KDE. The white dashed circles mark the scale radii $a$ and the dash-dotted lines are the Wd2 cluster are (see Sect.~\ref{sec:spat_dist}).}
\label{fig:kde}
\end{figure*}

To avoid confusion due to five groups and the larger number of stars, we divide the PMS stars into three different plots (bottom frames of Fig.~\ref{fig:kde}). The stars with an RV of $(-16.75 \pm 5.18)\,{\rm km}\,{\rm s}^{-1}$ (PMS1, yellow) appear to be distributed throughout the cluster region with the highest concentration of stars aligned with the the center of Wd2 along the MC--NC axis. The stars with RVs of $(-8.40 \pm 1.85)\,{\rm km}\,{\rm s}^{-1}$ (PMS2, green)) and $(3.10 \pm 2.46)\,{\rm km}\,{\rm s}^{-1}$ (PMS4, brown) tend to be located North of the cluster center. The remaining two RV groups with $(-3.57 \pm 2.00)\,{\rm km}\,{\rm s}^{-1}$ (PMS3, red) and $(-9.41 \pm 1.82)\,{\rm km}\,{\rm s}^{-1}$ (PMS5, black) are more concentrated toward the MC. Interestingly, the difference of the mean velocities of groups PMS2-PMS4 and PMS3-PMS5 is very similar with $(11.50 \pm 0.41)\,{\rm km}\,{\rm s}^{-1}$ and $(12.90 \pm 0.24)\,{\rm km}\,{\rm s}^{-1}$, respectively.

\subsubsection{The spatial location of the velocity groups}

The number of stars per velocity group is fairly low and to quantify the spatial correlation we use a 2D Kolmogorov-Smirnov (KS) test \citep{Hodges1958,Peacock1983,Fasano1987}. The null-hypothesis ($H_0$) we use is: two individual velocity groups follow the same spatial distribution. This means that if $H_0$ is true the $p$-value is larger than the significance level $\alpha$. We test $H_0$ against $\alpha = 5\%$ (confidence level: 95\%). In addition we also test their spatial locations against all Wd2 members of the HST photometric star catalog and the onse detected with MUSE. The resulting $p$-values are presented in Tab~\ref{tab:KS}. We also apply the 1D KS test we transform the two-dimensional location of the stars into one dimensional distribution by creating a cumulative distribution of the stars' distances to a reference point. The results of the 1D KS test highly vary with the choice of the reference point so we decided that the 1D KS test is not suited for our purposes. The KS-test results can be summarized the following:

\begin{itemize}
    \item HST vs. MUSE catalog: A $p$-value of 0.274 shows that their underlying spatial distribution is the same. This minimizes the chance of introducing correlations based on detection effects, such as completeness.
    \item Groups O1, B1, and B2: The 2D KS test confirms that these groups are spatially correlated to each other and to the full Wd2 MUSE catalog, in agreement with a mass segregated star cluster. We must note here that the $p$-value between the velocity groups B1 and B2 is only marginally significant.
    \item PMS2 -- PMS4: With $p = 0.045$ the correlation is only marginally significant. Yet, given the inspection of the KDE plot and that it is by far the highest $p$-value with respect to the other PMS groups let us conclude that these two groups are spatially correlated.
    \item PMS3 -- PMS5: The $p$-values suggest that these groups are correlated to each other but also to the O1, B1, B2, and PMS1 groups. Given their much different RV profile suggest that this is only the case because they are centered around the MC.
\end{itemize}

The 2D KS tests confirm our initial analysis about the spatial correlation of the PMS3 and PMS5 groups with the MC and the PMS2 and PMS4 groups with the NC. The PMS1 group is consistent with a group that follows the spatial distribution of the Wd2 cluster.

\begin{deluxetable*}{rrrrrrrrrrr}[htb]
	\tablecaption{The KS-test of the stellar spatial distributions \label{tab:KS}}
	\tablehead{\multicolumn{1}{c|}{}& \multicolumn{1}{c}{HST}& \multicolumn{1}{c}{MUSE}& \multicolumn{1}{c}{O1}& \multicolumn{1}{c}{B1}& \multicolumn{1}{c}{B2}& \multicolumn{1}{c}{PMS1}& \multicolumn{1}{c}{PMS2}& \multicolumn{1}{c}{PMS3}& \multicolumn{1}{c}{PMS4}& \multicolumn{1}{c}{PMS5}}
	\startdata
	\multicolumn{1}{r|}{HST} &  & \textbf{0.274} & \textbf{0.096} & \textbf{0.585} & 0.012 & \textbf{0.120} & 0.005 & \textbf{0.228} & 0.008 & \textbf{0.193} \\
	\multicolumn{1}{r|}{MUSE}& \textbf{0.274} &  & \textbf{0.108} & \textbf{0.889} & \textit{0.047} & \textbf{0.262} & 0.014 & \textbf{0.473} & 0.008 & \textbf{0.197} \\
	\multicolumn{1}{r|}{O1}  & \textbf{0.096} & \textbf{0.108} &  & \textbf{0.102} & \textbf{0.360} & 0.037 & 0.036 & \textbf{0.155} & 0.022 & \textbf{0.401} \\
	\multicolumn{1}{r|}{B1}  & \textbf{0.585} & \textbf{0.889} & \textbf{0.102} &  & \textbf{0.124} & \textbf{0.458} & \textbf{0.058} & \textbf{0.581} & 0.035 & \textbf{0.351} \\
	\multicolumn{1}{r|}{B2}  & 0.012 & \textit{0.047} & \textbf{0.360} & \textbf{0.124} &  & 0.021 & 0.003 & \textbf{0.143} & 0.002 & \textbf{0.119} \\
	\multicolumn{1}{r|}{PMS1}& \textbf{0.120} & \textbf{0.262} & 0.037 & \textbf{0.458} & 0.021 &  & 0.022 & \textbf{0.102} & 0.028 & \textbf{0.168} \\
	\multicolumn{1}{r|}{PMS2}& 0.005 & 0.014 & 0.036 & \textbf{0.058} & 0.003 & 0.022 &  & 0.014 & \textit{0.045} & 0.014 \\
	\multicolumn{1}{r|}{PMS3}& \textbf{0.228} & \textbf{0.473} & \textbf{0.155} & \textbf{0.581} & \textbf{0.143} & \textbf{0.102} & 0.014 &  & 0.007 & \textbf{0.242} \\
	\multicolumn{1}{r|}{PMS4}& 0.008 & 0.008 & 0.022 & 0.035 & 0.002 & 0.028 & \textit{0.045} & 0.007 &  & 0.027 \\
	\multicolumn{1}{r|}{PMS5}& \textbf{0.193} & \textbf{0.197} & \textbf{0.401} & \textbf{0.351} & \textbf{0.119} & \textbf{0.168} & 0.014 & \textbf{0.242} & 0.027 &  \\
	\enddata
	\tablecomments{This table shows the $p$-values of the 2D KS test for the spatial correlation between the individual velocity groups. To guide the reader's eye all $p$-values above 0.05 are marked in bold face, while the $p$-values that are marginally below 0.05 are marked in italic.}
\end{deluxetable*}

\subsubsection{Are the velocity groups a result of small number statistics?}
\label{sec:smal_number_stats}
Even though our sample consist of 117 cluster member stars, dividing them into 8 individual velocity groups leaves only a handful of stars per group (see Tab.~\ref{tab:RV_stars_comp}), which raises the question whether the individual groups are a result of small number statistics. In the following we estimate the likelihood that the five PMS RV groups are the result of a random occurrence by simulating 300 realizations of the PMS stellar RV distribution using Bayesian sampling. We test two different scenarios:

\begin{itemize}
    \item [1.] The true RV distribution only has one broad peak;
    \item [2.] The true RV distribution is similar to the distribution of the B-stars including a blue shifted component (representing the PMS1 peak).
\end{itemize}

For the first scenario, we sample the 300 different realizations using a likelihood distribution of one broad Gaussian with a mean velocity of $-2.12\,{\rm km}\,{\rm s}^{-1}$ and a velocity dispersion of $11.09\,{\rm km}\,{\rm s}^{-1}$, estimated from the cluster member RV distribution (see Fig.~\ref{fig:RV_dist}). For the second scenario we use a combination of three Gaussians, representing the PMS1, B1, and B2 groups (see Tab.~\ref{tab:RV_stars_comp}). For both scenarios the RV uncertainties are sampled from a likelihood distribution of the form $p(\sigma) \propto \sigma \cdot e^{\sfrac{-\sigma}{a}}$, which is a good (empirical) fit to the observed uncertainties (see Fig.~\ref{fig:err_dist}).

We then try to recover the true RV distribution of each scenario, as well as five RV groups. We use the same priors and technique as we used for the real data. To analyze the results we compare the number of MCMC runs that reach convergence\footnote{We consider only those that converged to one value for each parameter and each parameter stays well within set boundaries. The latter ensures physically meaningful results (e.g., no negative amplitudes).}. For the first scenario in only 16.0\% (48 out of 300 realizations) we find five peaks while in 80.3\% (241 out of 300 realizations) the underlying true distribution can be recovered. For scenario two in 26.0\% (75 out of 300 realizations) we find five peaks while in 75.0\% (225 out of 300 realizations) the underlying distribution can be recovered. These results, in combination with the correlation of the spatial location of the stars and their membership to certain RV groups, suggest that it is unlikely that small number statistics are the reason for the five groups. Although the five RV peaks are the most probable result, a second, independent dataset like higher resolution spectroscopy or high-precision astrometry (see Sect.~\ref{sec:Gaia}) may provide an independent confirmation.

\subsection{The dynamical state}
\label{sec:duy_state}

To determine whether this cluster has the chance of overcoming the ``infant mortality'' we will make an assessment of its dynamical state by estimating its dynamical mass $M_{\rm dyn}$, the viral radius $r_{\rm vir}$, and the dynamical time $t_{\rm dyn}$ or crossing time \citep[detailed derivations and discussions of these parameters can be found in][and references therein]{Spitzer1987,Fleck2006,PortegiesZwart2010,Krumholz2020,Adamo2020}.

The dynamical mass is defined as follows:

\begin{equation}
\label{eq:M_dyn}
    M_{\rm dyn} = \eta \left(\frac{\sigma^2 r_{\rm hm}}{G}\right),
\end{equation}

where $\sigma$ is the 1D velocity dispersion, $G$ the gravitational constant, $r_{\rm hm}$ the half-mass radius, and $\eta$ is a dimensionless parameter to link observational accessible parameters with theory and is typically $\eta = 9.75$ for clusters with $\gamma > 4$ (see Tab.~\ref{tab:spat_dist} for Wd2 parameters). For this value, the virial radius is $r_{\rm vir} = 1.625 \cdot r_{\rm hm}$.

The typical assumption is that massive star clusters are spherically symmetric, which is not the case for Wd2. Therefore, we decided to analyze the following cases: 1) the MC and NC are separate clusters; 2) the MC and NC are located at the same position in space with $r_{\rm hm,Wd2} = (r_{\rm hm,MC} + r_{\rm hm,NC}) / 2$; 3) the half mass radius of Wd2 incorporates both the MC and the NC, hence $r_{\rm hm,Wd2} = r_{\rm hm,NC} + r_{\rm hm,MC} + d({\rm MC},{\rm NC})$; and 4) a thought experiment on which is the minimal necessary half-mass radius for a bound spherical cluster with the photometric mass of Wd2. The results are:

\begin{itemize}
    \item[1.] Due to the high degree of mass segregation we use the mean velocity dispersion of the PMS velocity groups PMS3,PMS5: $(2.16 \pm 0.49)\,{\rm km}\,{\rm s}^{-1}$, and PMS2,PMS4: $(1.91 \pm 0.26)\,{\rm km}\,{\rm s}^{-1}$, for the MC and NC, respectively. These yield $M_{\rm dyn,MC} = (1.9 \pm 0.5)\cdot10^3\,{\rm M}_\odot$ and $M_{\rm dyn,NC} = (3.3 \pm 1.4)\cdot10^3\,{\rm M}_\odot$.
    \item[2.] The assumed half-mass radius of Wd2 is $r_{\rm hm,Wd2} = (0.13\pm0.04)\,{\rm pc}$. The velocity dispersion, $\sigma_{\rm Wd2} = (11.09 \pm 1.36)\,{\rm km}\,{\rm s}^{-1}$, is estimated from the cluster member velocity distribution (see Fig.~\ref{fig:RV_dist}). This yields a dynamical mass of $M_{\rm dyn,Wd2} = (7.5 \pm 1.9)\cdot10^4\,{\rm M}_\odot$.
    \item[3.] The assumed half-mass radius of Wd2 incorporating the MC and NC is $r_{\rm hm,Wd2} = (1.57\pm0.01)\,{\rm pc}$, which leads to a dynamical mass of $M_{\rm dyn,Wd2} = (4.4 \pm 1.1)\cdot10^5\,{\rm M}_\odot$ (with $\sigma_{\rm Wd2}$ as in 2.).
    \item[4.] We assume $M_{\rm dyn} = M_{\rm phot} = (3.7 \pm 0.8)\cdot10^4\,{\rm M}_\odot$ \citep{Zeidler2017}. With $\sigma_{\rm Wd2}$ as in 2., the half mass radius is $r_{\rm hm} = (0.13\pm0.04)\,{\rm pc}$.
\end{itemize}

The dynamical time, or crossing time, is the time that a star needs to cross the cluster system. It indicates how long a system needs to establish or re-establish dynamical equilibrium. It is defined as:

\begin{equation}
\label{eq:t_dyn}
    t_{\rm dyn} = \sqrt{\frac{r_{\rm vir}^3}{GM_{\rm phot}}}.
\end{equation}

For the results of the four cases the dynamical time yields: 1.) $t_{\rm dyn,MC} = 0.3\,{\rm Myr}$ and $t_{\rm dyn,NC} = 1.6\,{\rm Myr}$; 2.) $t_{\rm dyn,Wd2} = 0.034\,{\rm Myr}$, 3.) $t_{\rm dyn,Wd2} = 4.64\,{\rm Myr}$, and 4.) $t_{\rm dyn,Wd2} = 0.11\,{\rm Myr}$.

It becomes clear that the dynamical mass and time highly depend on the structure of the underlying system and the assumptions made to determine these parameters, which we will discuss in detail in Sect.~\ref{sec:discussion}.

\section{The \textit{G\lowercase{aia}} DR2}
\label{sec:Gaia}
The Wd2 cluster and its parental \ion{H}{2} region RCW49 is being observed by the \textit{Gaia} satellite and the already collected data is part of the data release 2 \citep[DR2,][]{GaiaCollaboration2016,GaiaCollaboration2018}. Its location in the Carina-Sagittarius spiral arm and the extinction and crowding impose limitations to the DR2 accuracy. Hence, many cluster member parameters, such as stellar velocities, and parallaxes are still poorly constrained\footnote{For example, only three stars of the HST photometric catalog have \textit{Gaia} RVs (Paper 1)}. 

Nevertheless, we analyze the existing \textit{Gaia} stellar proper motions (pms) using priors based on the knowledge we gained from the HST and MUSE data. We cross-correlate the HST and the \textit{Gaia} catalogs. Of the 20,482 point sources in the HST catalog, 1239 are included in the \textit{Gaia} DR2, of which 471 are cluster member stars based on the HST CMD selection \citep{Zeidler2015,Sabbi2020}. To select stars with a clean astrometric solution we use the following magnitude based limits as suggested by \citet{Lindegren2018}:

\begin{equation}
\label{eq:clean_astrometric_solution_gaia}
    u < 1.2 \cdot \max{\left(1, \exp{\left(-0.2\cdot\left(G-19.5\right)\right)}\right)},
\end{equation}

where $G$ is the \textit{Gaia} $G$-band magnitude and $u = \left(\chi^2 / \nu \right)^{\sfrac{1}{2}}$. $\chi$ is the astrometric goodness-of-fit in the ``along-scan'' direction and $\nu$ is the adjoined number of good observations. This leaves us with 282 cluster members, of which 85 also have MUSE RVs. We use the 282 cluster members to calculate the systemic pms in R.A. and Dec. $\mu_{\alpha \ast,{\rm sys}} = -5.17\,{\rm mas}\,{\rm yr}^{-1}$\footnote{$\mu_{\alpha \ast}$ is the deprojected, declination corrected pm in R.A.: $\mu_\alpha \cdot \cos{\delta}$} and $\mu_{\delta,{\rm sys}} = 3.00\,{\rm mas}\,{\rm yr}^{-1}$, which is $\mu_{\alpha \ast,{\rm sys}} = -101.9\,{\rm km}\,{\rm s}^{-1}$ and $\mu_{\delta,{\rm sys}} = 59.1\,{\rm km}\,{\rm s}^{-1}$ at the distance of Wd2 (4.16\,kpc). As for the cluster RVs (Sect.~\ref{sec:RV}) we subtract the systemic velocities from the cluster members throughout the rest of this work unless stated otherwise.

\begin{figure}[htb]
\plotone{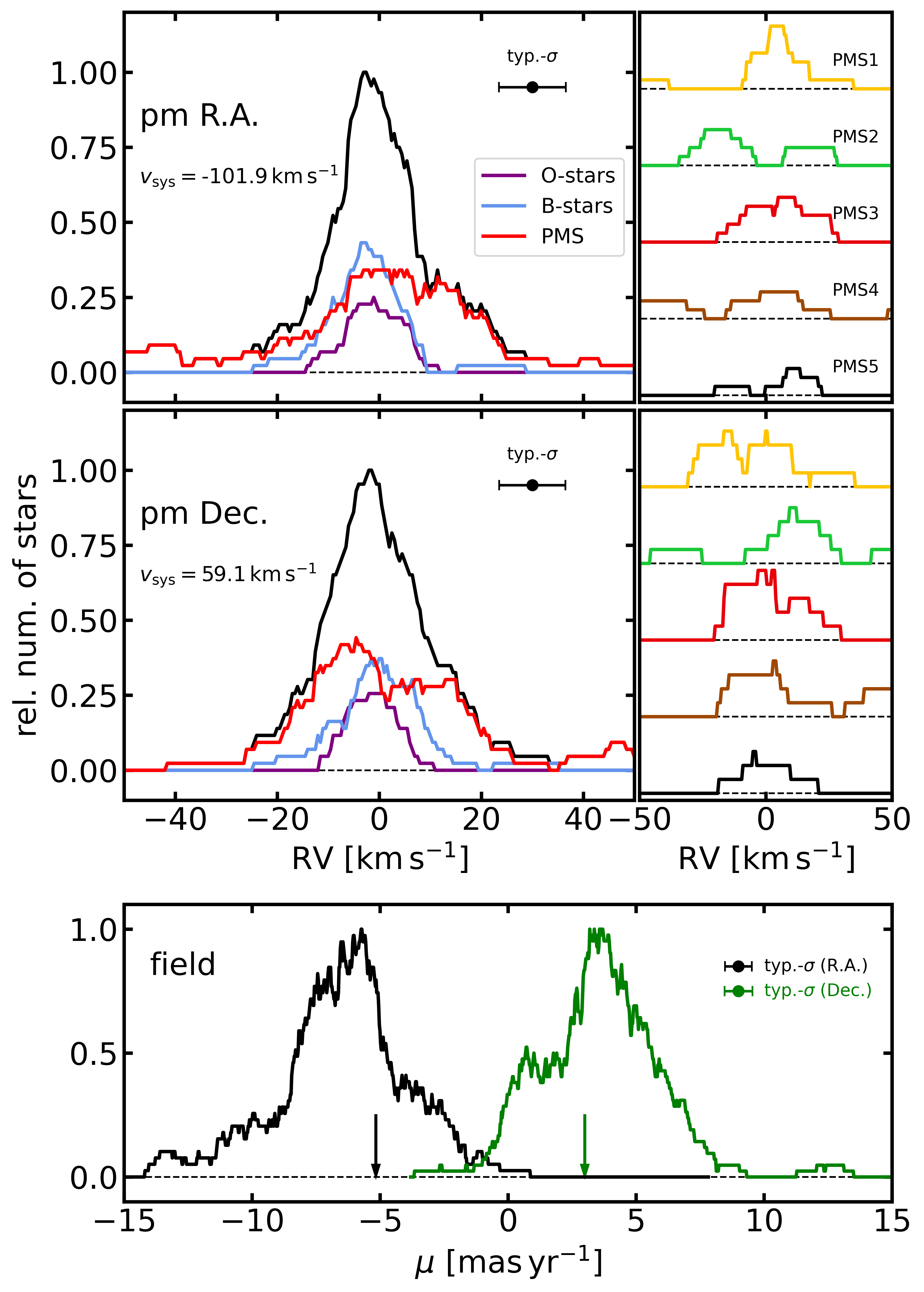}
\caption{The \textit{Gaia} pm distributions of the stars toward the Wd2 cluster. The left two plots of the top panel show the pm distributions in R.A. and Dec. for all cluster member stars (black), O-stars (purple), B-stars (blue), and PMS stars (red) computed with the same technique as RV distribution (see Fig.~\ref{fig:RV_dist} and \ref{fig:RV_dist_grouped}). On the right we show the pm distribution of the five RV groups (see Fig.~\ref{fig:RV_dist_grouped}). The bottom panel shows the pm distributions of the foreground field stars (R.A. in black and Dec. in green). The arrows indicate the systemic pms of the Wd2 stars.}
\label{fig:pm_dist}
\end{figure}

To analyze the pm distributions we only use the 85 stars that have MUSE RVs. This allows us to differentiate between O-stars, B-stars, and PMS stars. For the pm distributions we use the same method as for the RVs (Sect.~\ref{sec:RV_stars}). The typical uncertainties are $\sigma \mu_{\alpha \ast} = 0.332\,{\rm mas}\,{\rm yr}^{-1}$ ($6.55\,{\rm km}\,{\rm s}^{-1}$) and $\sigma \mu_{\delta} = 0.329\,{\rm mas}\,{\rm yr}^{-1}$ ($6.49\,{\rm km}\,{\rm s}^{-1}$) for cluster members, and $\sigma \mu_{\alpha \ast} = 0.524\,{\rm mas}\,{\rm yr}^{-1}$ and $\sigma \mu_{\delta} = 0.542\,{\rm mas}\,{\rm yr}^{-1}$ for field stars. We show the field star distributions in the bottom panel of Fig.~\ref{fig:pm_dist} (R.A. in black and Dec. in green). The arrows indicate the respective cluster member systemic pms. Similar to the RVs (see Fig.~\ref{fig:RV_dist}) the velocity space of the field stars and the cluster members overlap. This means that also pms are not suitable to improve the cluster member selection.

The pm profiles of the O-stars, B-stars, and PMS stars (top panels of Fig.~\ref{fig:pm_dist}) indicate a similar distribution as the RV profile. The O-stars and B-stars are centered around the $0\,{\rm km}\,{\rm s}^{-1}$ with the B-stars showing a slightly broader velocity profile. The distribution of the PMS is much broader covering more than $20\,{\rm km}\,{\rm s}^{-1}$. While they are also centered around the systemic pm in declination they appear to be slightly offset in R.A.. In both pm directions the PMS stars show two peaks at $\sigma \mu_{\alpha \ast} \approx 0\,{\rm km}\,{\rm s}^{-1}$ and $\sigma \mu_{\alpha \ast} \approx 12\,{\rm km}\,{\rm s}^{-1}$ and $\sigma \mu_{\delta} \approx -5\,{\rm km}\,{\rm s}^{-1}$ and $\sigma \mu_{\delta} \approx 8\,{\rm km}\,{\rm s}^{-1}$. The fairly high pm uncertainties (in comparison to the RVs) do not allow to resolve the five individual velocity groups detected with the RVs, assuming they have a similar separation and dispersion in pm direction. Additionally, the \textit{Gaia} DR2 data is less deep than the MUSE dataset, which may lead to a similar effect we saw in Paper 1 for the shallower MUSE sample. Here also only two RV peaks where detected ($\sim 17\,{\rm km}\,{\rm s}^{-1}$ apart). In the right top panels of  Fig.~\ref{fig:pm_dist} we show the pm distribution of the five PMS RV groups. Their distributions indicate marginal relative shifts, yet the relatively large uncertainties and the low numbers (not all RV stars have pms) do not allow for a confirmation of the velocity groups. Future \textit{Gaia} data releases will provide the necessary precision.

\section{High velocity runaway candidates}
\label{sec:runaways}
High stellar densities in YSCs (either naturally formed or through rapid mass-segregation), binaries, and higher-order systems increase the probability for close encounters within the cluster \citep[e.g., ][and reference therein]{PortegiesZwart2010}. These dynamical interactions and supernova explosions (if one binary component explodes) can give a star a ``kick'' making it a runaway star. Runaway events are believed to be the main source of populating the field with these massive objects \citep{Blaauw1961,PortegiesZwart2010}. In the MW, sources with a velocity $>30\,{\rm km}\,{\rm s}^{-1}$ relative to the local standard of rest are considered as runaway stars \citep{Hoogerwerf2001}. Recent studies found that a significant number of massive O and B-stars may have been ejected from YSCs including Wd2, NGC3603, and R136 \cite{Roman-Lopes2011,Lennon2018,Drew2018,Drew2019}.

\begin{figure}[htb]
\plotone{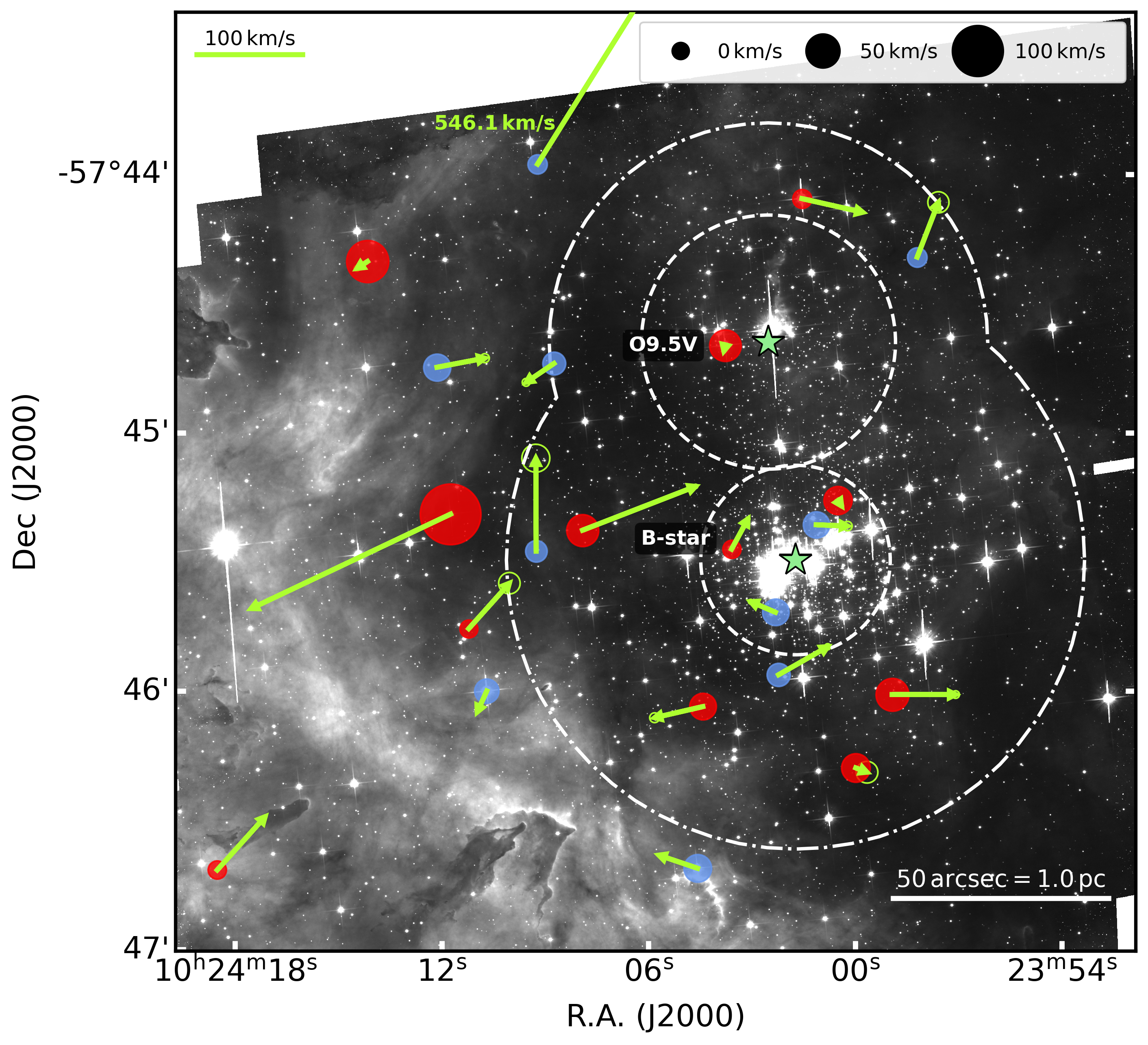}
\caption{The stars with an absolute peculiar velocity exceeding $30\,{\rm km}\,{\rm s}^{-1}$. The green arrows indicate the value and direction of the proper motions. At the arrow tip we show in green the pm error ellipses. The RVs are shown as red and blue circles depending if the stars move away from us or toward us. The circle size is an indicator for the RV value. For displaying proposes we cut the arrow of the star with a peculiar velocity of $546.1\,{\rm km}\,{\rm s}^{-1}$.}
\label{fig:runaways}
\end{figure}

In both, the MUSE RV distribution and the Gaia DR2 pm distributions (see Fig.~\ref{fig:RV_dist_grouped} and \ref{fig:pm_dist}) we see bona-fide cluster member stars with velocities exceeding $\pm30\,{\rm km}\,{\rm s}^{-1}$. To analyze these runaway candidates we calculate their peculiar velocities based on the three velocity components (RV, $\mu_{\alpha \ast}$, $\mu_{\delta}$). We only consider those sources, whose total pm uncertainty do not exceed 50\% prior to the systemic velocity subtraction, which ensures that a proper direction of the stars' motion can be determined. The pm uncertainty ellipse is determined following eq.~(9) in \citet{Lindegren2016} and eq.~(B.2) in \citet{Lindegren2018}. Since we are only interested in the the relative motion of the stars, we do not consider any systematic offsets in the DR2 pms and parallaxes as they are described in \citet{Lindegren2018}. In total we find 22 stars that fulfill the above criteria, of which one is an O9.5V star\footnote{Based on the spectral type determined by \citet{VargasAlvarez2013} and confirmed in Paper 1.} (ID: 10198) and one is a B-type star (ID: 10048). The parameters of all 22 sources are listed in Tab.~\ref{tab:runaways}. In Fig.~\ref{fig:runaways} we show the location of the runaway candidates. The green arrows point in the direction of the stars movement in pm space, while the length of the vector represents the velocity. The ellipse at each vector's tip represents the pm error ellipse. The size of each point represents the magnitude of the RV (blue/red points indicate a relative RV toward/away from the Sun).

The majority of runaway candidates show peculiar velocities in the range of 30--$100\,{\rm km}\,{\rm s}^{-1}$. Only three stars exceed this range: ID-13587\footnote{r.a.~$=10^\mathrm{h}24^\mathrm{m}07.91^\mathrm{s}$, dec.~$=-57^\circ45{}^\prime22.69{}^{\prime\prime}$} with $123.2 \pm 4.2\,{\rm km}\,{\rm s}^{-1}$, ID-16306\footnote{r.a.~$=10^\mathrm{h}24^\mathrm{m}11.74^\mathrm{s}$, dec.~$=-57^\circ45{}^\prime18.94{}^{\prime\prime}$} with $245.8 \pm 2.3\,{\rm km}\,{\rm s}^{-1}$, and ID-14542\footnote{r.a.~$=10^\mathrm{h}24^\mathrm{m}09.22^\mathrm{s}$, dec.~$=-57^\circ43{}^\prime57.67{}^{\prime\prime}$} with $546.1 \pm 5.3\,{\rm km}\,{\rm s}^{-1}$.

\section{Discussion}
\label{sec:discussion}

In the following we will discuss the results presented in this work on the internal dynamics of Wd2.

\citet{Furukawa2009}, \citet{Ohama2010} and \citet{Fukui2016} argued, based on the results of NANTEN2 CO sub-millimeter observations, that cloud-cloud collision of two CO clouds at $4\,{\rm km}\,{\rm s}^{-1}$ and $16\,{\rm km}\,{\rm s}^{-1}$ may have triggered the formation of Wd2. Our RV analysis of the \ion{H}{2} region RCW49 surrounding Wd2 shows that its mean RV of $15.9\,{\rm km}\,{\rm s}^{-1}$ is in agreement with their conclusion. Furthermore, is the cavity, created by the ionizing fluxes of the many OB stars, expanding at a rate of $\sim 15\,{\rm km}\,{\rm s}^{-1}$ (see Fig.~\ref{fig:EBV_RVgas}). We must note that projection effects and the limited survey area may have an influence on that number, which explains the asymmetric RV distribution between the bottom and top 10\% of the velocity distributing ($\le -5.54\,{\rm km}\,{\rm s}^{-1}$ and $\ge 9.95\,{\rm km}\,{\rm s}^{-1}$, see Fig.~\ref{fig:EBV_RVgas_analysis}). Hence, the $\sim 15\,{\rm km}\,{\rm s}^{-1}$ should be considered as a lower limit. Nevertheless, the expansion rate of $\sim 7{\rm -}10\,{\rm km}\,{\rm s}^{-1}$ is comparable with studies of other \ion{H}{2} regions, such as N44 \citep[$\sim 6 {\rm -} 11\,{\rm km}\,{\rm s}^{-1}$,][]{Naze2002,McLeod2019}, N11 and N180 \citep[$\sim 10\,{\rm km}\,{\rm s}^{-1}$ and $10 {\rm -} 20\,{\rm km}\,{\rm s}^{-1}$, respectively,][]{Naze2001}, and other Magellanic Clouds, Milky Way, and extra galactic \ion{H}{2} regions \citep[e.g.,][]{Murray2009,Mesa-Delgado2010,McLeod2020}. These numbers are also supported by a variety of  numerical models and simulations \citep{Osterbrock1989,Bertoldi1990,Fujii2016,Haid2018}. The stellar and gas velocities are uncorrelated. We conclude that the \ion{H}{2} region is dominated by feedback processes, such as stellar winds and radiation pressure \citep[see e.g.,][]{Dale2015a} and has lost the imprint of the original cloud collapse.

In Paper 1 we demonstrated that cluster member stars show two distinct RV groups. The use of the short exposures only led to a smaller sample of stars with a cutoff at higher masses, which led to a bias toward more massive stars. Combining the short and long exposures and using the full capacity of \MUSEpack~ allows us to conduct a more sophisticated study of the stellar RV distribution. The three main results of this analysis are: 1) stars of different masses show different RV distributions 2) the lower the stellar mass, the higher the velocity dispersion, and 3) the low-mass PMS stars show five distinct, spatially correlated RV groups. The distributions of the O and B stars (one and two peaks, see Fig.~\ref{fig:RV_dist_grouped}) are in agreement with the two RV peaks detected in \citetalias{Zeidler2018}. The overall smaller RV dispersion of the OB-stars is in good agreement with a highly mass-segregated star cluster. Two-body relaxation drives star clusters toward energy equipartition \citep[$m_i v_i^2 = \rm const. $, e.g.,][]{Spitzer1969, Parker2016}, which impacts high-mass stars faster and stronger. We also must note that, given the young age and the EFF profile being the best-fitted mass distribution it is almost impossible that Wd2 has reached energy equipartition. The low-mass PMS stars do not only show an overall higher velocity dispersion, they also belong to five distinct velocity groups (see Fig.~\ref{fig:RV_dist_grouped}). While PMS groups 2 -- 5 have very similar velocity dispersions ($\sim 2\,{\rm km}\,{\rm s}^{-1}$, see Tab.~\ref{tab:RV_stars_comp}), the velocity dispersion of PMS group 1 is with $5.18\,{\rm km}\,{\rm s}^{-1}$ much higher. The analysis of the spatial location of the stars in each velocity reveals a correlation. Always two groups (PMS3,PMS5 and PMS2,PMS4) coincide with the MC and NC. The stars of PMS group 1 are distributed throughout the cluster region in a halo-like structure with a higher concentration toward the center of Wd2. Observations and theoretical studies show that star and star cluster formation is a hierarchical process and that YSCs form through merging of smaller sub cluster \citep[e.g.,][]{McMillan2007a,Sabbi2007,Fujii2012,Banerjee2015,Fujii2016}. Given the age of Wd2 (1--2\,Myr) we conclude that the individual velocity groups are a remnant of the formation process of the MC and NC. While the feedback from the OB stars has destroyed this imprint in the \ion{H}{2} region the stars are much less affected by feedback processes. The fact that always two groups are co-located with each clump may either suggest that these clumps have been more sub structured or it is a residual from the possible cloud-cloud collision that initiated the formation of Wd2. The latter is supported by the fact the mean velocity differences of PMS2,PMS4 and PMS3,PMS5 are $(11.50 \pm 0.41)\,{\rm km}\,{\rm s}^{-1}$ and $(12.90 \pm 0.24)\,{\rm km}\,{\rm s}^{-1}$, respectively, which is in agreement with the velocity difference of the two CO clouds \citep[$4\,{\rm km}\,{\rm s}^{-1}$ and $16\,{\rm km}\,{\rm s}^{-1}$,][]{Furukawa2009}. Although the two clumps of Wd2 are coeval \citep{Zeidler2015}, its structure has similarities with the observations of a highly sub-structured ONC with several kinematically different stellar groups at different ages \citep{Zari2019a}.

Star cluster populations throughout the Universe show a drop in number for YSCs (typically around $<10$\,Myr) compared to the older cluster population ($>100$\,Myr). This is often referred to as ``infant mortality'' \citep[e.g.,][]{Lada2003,Goodwin2006,PortegiesZwart2010} and means that there is a process that destroys the majority of YSCs ($\sim90\%$) within the first few 10\,Myr of their life. This typically happens because these YSCs are not massive enough to overcome internal (e.g., supernova explosions, rapid gas expulsion) and external (e.g., collisions and close encounters with GMC in the Galactic disk, changes in the external tidal field) evolution effects. In this work we solely focus on the internal processes and a detailed discussion on external factors can be found in e.g., \citet{Krumholz2019}. The absolute minimal mass necessary to keep a self-gravitating system bound is: $M_{\rm dyn} = M_{\rm phot}$. The non-spherical substructured nature of Wd2 makes this comparison challenging and we introduced four different cases. In case 1. we consider the two clumps as individual clusters and the resulting dynamical masses, $M_{\rm dyn,MC} = (1.9 \pm 0.5)\cdot10^3\,{\rm M}_\odot$ and $M_{\rm dyn,NC} = (3.3 \pm 1.4)\cdot10^3\,{\rm M}_\odot$ are smaller than the individual photometric masses \citep[$M_{\rm phot,MC} = (2.8 \pm 0.6)\cdot10^4\,{\rm M}_\odot$ and $M_{\rm phot,NC} = (4.2 \pm 1.3)\cdot10^3\,{\rm M}_\odot$,][]{Zeidler2017}. This is an unrealistic scenario and the close proximity of the two clumps suggests that they will merge in the near future supporting hierarchical formation of YSCs \citep[e.g.,][]{Fujii2012}. Therefore, we must analyze the cluster system as a whole. In case 2. we assume the MC and NC are located at the same position with $r_{\rm hm,Wd2} = (r_{\rm hm,MC} + r_{\rm hm,NC}) / 2$ and in case 3. the half mass radius incorporates both the MC and the NC ($r_{\rm hm,Wd2} = r_{\rm hm,MC} + r_{\rm hm,NC} + d({\rm MC},{\rm NC})$). Both cases lead to a dynamical mass ($M_{\rm dyn,Wd2} = (7.5 \pm 1.9)\cdot10^4\,{\rm M}_\odot$ and $M_{\rm dyn,Wd2} = (4.4 \pm 1.1)\cdot10^5\,{\rm M}_\odot$) that highly exceeds the Wd2 photometric mass. Even without any external perturbations this suggests that Wd2 will disperse in the future. This is supported by an unreasonably small half-mass radius ($r_{\rm hm} = (0.13\pm0.04)\,{\rm pc}$, in comparison the mass surface density, see Fig.~\ref{fig:spat_dist}) that is necessary for case 4, $M_{\rm dyn} = M_{\rm phot}$.

Next we discuss the dynamical time scale. The ratio of the cluster's age with its dynamical time ($\Pi = {\rm age}/{t_{\rm dyn}}$) should be large if a system is bound, while for unbound systems it is expected to be small. A cut can be defined at $\Pi \sim 1$--3 but has to be used with caution when the ratio is close to this, somewhat arbitrary cut \citep{Pfalzner2009,Adamo2020}. For an age range of 1--2\,Myr for Wd2 \citep[e.g.,][]{Zeidler2015,Sabbi2020} this ratio is $\Pi = 29$--58, $\Pi = 0.22$--0.43, and $\Pi = 9$--18 for cases 2., 3., and 4., respectively. $\Pi = 9$--18 (case 4.) agrees with the value of $\Pi = 8.48$ shown in Tab.~2 of \citet{PortegiesZwart2010}, yet this value is based on a distance of 2.8\,kpc for Wd2 \citep{Pfalzner2009}. The somewhat high variation of this ratio does not allow a conclusive determination of whether Wd2 is bound. Yet, the analyses of the dynamical mass and its location in the Galactic disk lead to the conclusion that Wd2 will eventually dissolve. In addition to external perturbations, the many OB-stars in the cluster center will go supernovae, ejecting the remaining gas within the cluster. This will shock the cluster's gravitational potential leading to an almost instantaneous expansion \citep[see e.g.,][]{Goodwin2006}, which will accelerate this dispersion.

While not all (massive) stars from in clusters \citep[e.g.,][]{DeWit2005,Ward2018}, high-velocity runaway stars are the main source for massive O and B field stars \citep[e.g.,][]{Blaauw1961,PortegiesZwart2010,Fujii2011,Oh2016}. In the larger vicinity ($1.5 \times 1.5\,{\rm deg}^2$) around Wd2, \citet{Drew2018} detected 8--11 early O and Wolf-Rayet stars that are Wd2 runaway candidates. In this work the combination of \textit{Gaia} DR2 proper motions and the high accuracy of the MUSE RVs allowed us to detect high-velocity runaway stars inside the cluster region. The majority of runaway candidates show peculiar velocities in the range of 30--$100\,{\rm km}\,{\rm s}^{-1}$ but three stars exceed this range: ID-13587 with $123.2 \pm 4.2\,{\rm km}\,{\rm s}^{-1}$, ID-16306 with $245.8 \pm 2.3\,{\rm km}\,{\rm s}^{-1}$, and ID-14542 with $546.1 \pm 5.3\,{\rm km}\,{\rm s}^{-1}$. We do not detect any preferred ejection direction (see Fig.~\ref{fig:runaways}), which is in agreement that these high stellar velocities are obtained through two-body interactions in the dense cluster center. Although the favored scenario for these ``kicks'' are supernovae explosions of the binary component \citep[e.g.,][]{Hoogerwerf2001}, we concur with \citet{Drew2018} that this scenario is very unlikely because Wd2 is too young. Especially the massive runaway stars are important to understand the initial mass function (IMF). While studies show that some YSCs, including Wd2 \citep{Zeidler2017}, show a top-heavy present day MF, YSCs that show a canonical IMF may have had a top-heavy IMF before the early ejections of some of their most massive members.

\section{Summary and Conclusions}
\label{sec:summary}

The young age, the close proximity, and its spatial sub-structure make Wd2 an interesting target to study how YSCs form and evolve during their first few million years. In this third paper of this series, we used the unique capabilities of HST photometry, \textit{Gaia} pms, and VLT/MUSE RVs to analyze the internal kinematic structure of Wd2 with the following results:

\begin{itemize}
    \item The current, already super-virial state of Wd2, the fact that the first supernovae are yet to happen, and its location in the MW disk, which make Wd2 prone to interact or even collide with other GMCs, led to the conclusion that the cluster is not massive enough to remain gravitationally bound.
    \item The cluster velocity dispersion increases with decreasing stellar mass, as expected from a highly mass-segregated cluster.
    \item The low-mass PMS stars have five distinct and statistically significant velocity groups.
    \item Always two velocity groups are part of each of the two clumps (MC and NC), while the fifth group is a halo-like structure, in agreement with the formation of star clusters through mergers.
    \item We detected 22 runaway candidates that may be kicked out of the cluster due to two-body interactions caused by the high stellar density in the cluster center.
     \item The \ion{H}{2} region that surrounds Wd2 is expanding, driven by the radiation pressure and FUV flux of the many cluster center OB stars. Any imprint of the original cloud collapse has been destroyed.
\end{itemize}

Although this analyses already provides a multi-dimensional picture, future data releases of the \textit{Gaia} mission and multi epoch HST observations to accurately measure stellar proper motions will allow truly 3D kinematic studies.

\acknowledgments
We thank S. Kamann for his continuous support in using \texttt{Pampelmuse}. We also thank N. Miles for his support with the parallelization of python scripts, specifically using \texttt{dask}, and T. Morishita for his MCMC support. We thank P. Sonnentrucker for fruitful and interesting scientific discussions. We also thank the anonymous referee for their help to improve this paper.

P.Z. acknowledges support by the Forschungs- stipendium (ZE 1159/1-1) of the German Research Foundation.

This work is partly supported by NASA through the NASA Hubble Fellowship grant HF2-26555 (AFM).

This work has made use of data from the European Space Agency (ESA) mission {\it Gaia} (\url{https://www.cosmos.esa.int/gaia}), processed by the {\it Gaia} Data Processing and Analysis Consortium (DPAC, \url{https://www.cosmos.esa.int/web/gaia/dpac/consortium}). Funding for the DPAC has been provided by national institutions, in particular the institutions participating in the {\it Gaia} Multilateral Agreement.

These observations are associated with program \#14807. Support for program \#14807 was provided by NASA through a grant from the Space Telescope Science Institute. This work is based on observations obtained with the NASA/ESA \textit{Hubble} Space Telescope, at the Space Telescope Science Institute, which is operated by the Association of Universities for Research in Astronomy, Inc., under NASA contract NAS 5-26555.

\vspace{5mm}
\facilities{HST(WFC3,ACS), VLT(MUSE), \textit{Gaia}}


\software{Astropy \citep{Astropy2018}, Dask \citep{DaskDevelopmentTeam2016}, ESORex \citep{Freudling2013}, pyspeckit \citep{Ginsburg2011}, Matplotlib \citep{Hunter2007}, MUSEpack \citep{Zeidler2019}, MUSE pipeline \citep[v.2.8.1][]{Weilbacher2012,Weilbacher2015}, PampleMuse \citep{Kamann2013,Kamann2016}, pPXF \citep{Cappellari2004,Cappellari2017}}



\appendix

\section{Uncertainty propagation for the EFF profile}
\label{sec:error_prop}
\begin{equation}
\label{eq:EFF_mass}
    m(r) = \frac{2\pi\Sigma_0a^2}{\gamma - 2}\cdot\left[1-\left(1+\frac{r^2}{a^2}\right)^{1-\sfrac{\gamma}{2}}\right]
\end{equation}

\begin{equation}
\label{eq:EFF_mass_tot}
    M_{\rm tot} = \frac{2\pi\Sigma_0a^2}{\gamma - 2}
\end{equation}

\begin{equation}
\label{eq:sEFF_mass_tot}
    \delta M_{\rm tot} = \frac{2\pi a}{\gamma-2}\sqrt{\delta \Sigma_0^2 + \Sigma_0^2\left(4\delta a^2 + \frac{a^2}{\left(\gamma-2\right)^2}\delta \gamma^2\right)}
\end{equation}

\begin{equation}
\label{eq:sEFF_rc}
    \delta r_c = \sqrt{\left(2^{\sfrac{\gamma}{2}}-1\right)\delta a^2 + \frac{2^\gamma a^2 \log{2}}{16\left(2^{\sfrac{\gamma}{2}}-1\right)} \delta \gamma^2}
\end{equation}

\begin{equation}
\label{eq:sEFF_rhm}
    \delta r_{\rm hm} = \sqrt{\left(2^\frac{-2}{2-\gamma} -1\right)^2 \delta a^2 + \frac{a^2\log{0.5}^2}{\left(2-\gamma\right)^4 \cdot \left(2^\frac{-2}{2-\gamma} -1\right) \cdot \left(2^\frac{-4}{2-\gamma} -1\right)}\delta \gamma^2}
\end{equation}

\section{Plots}
\label{sec:plots}

\begin{figure}[htb]
\plotone{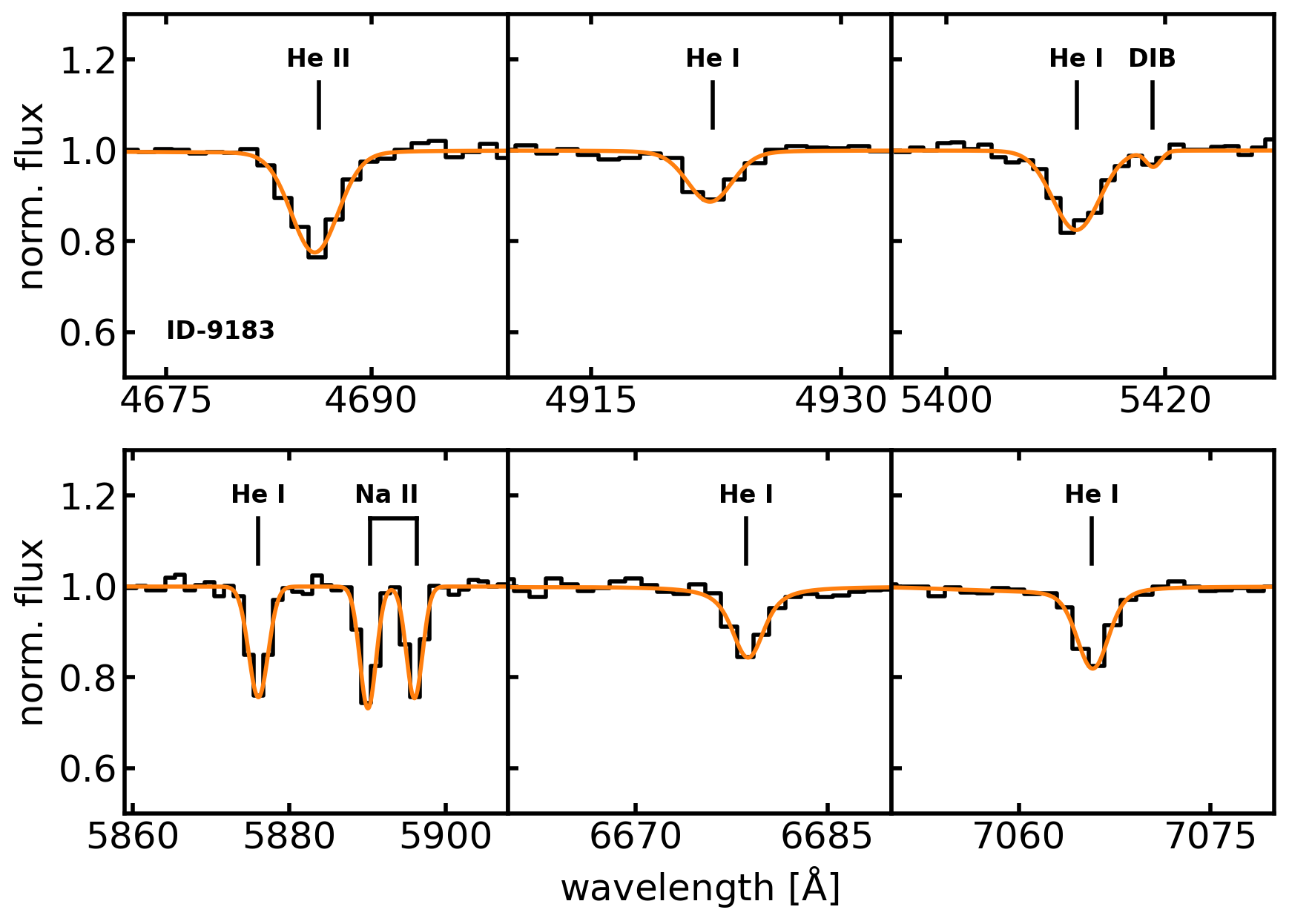}
\caption{The fitted spectrum for star 9183. The extracted and rectified stellar spectrum in the regions of strong \ion{He}{1} and \ion{He}{2} lines is shown in black and with the best spectral fit in orange.}
\label{fig:spec_fit_HeIHeII_9183}
\end{figure}

\begin{figure}[htb]
\plotone{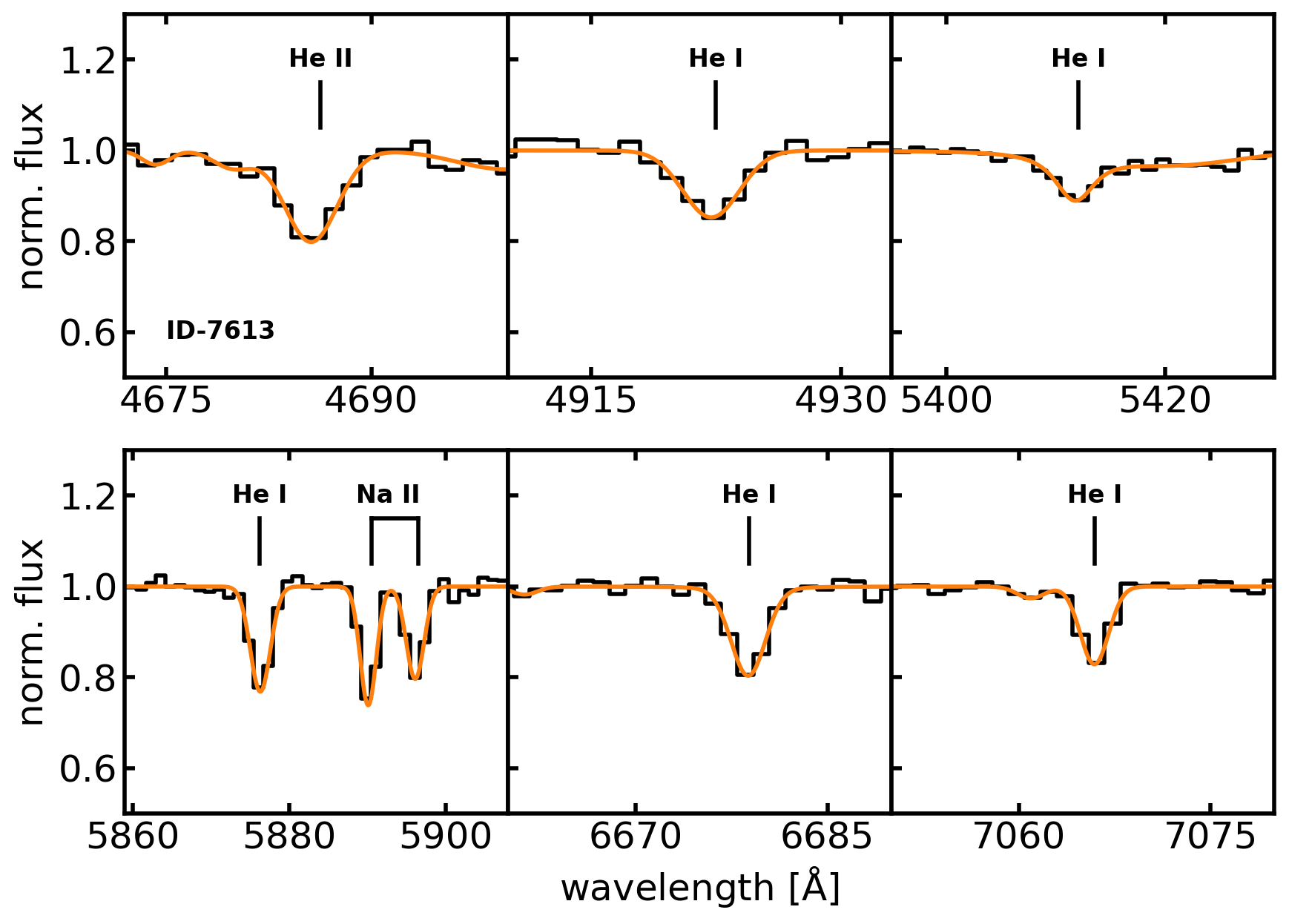}
\caption{The fitted spectrum for star 7613. The plot is similar to Fig.~\ref{fig:spec_fit_HeIHeII_9183}.}
\label{fig:spec_fit_HeIHeII_7613}
\end{figure}

\begin{figure}[htb]
\plotone{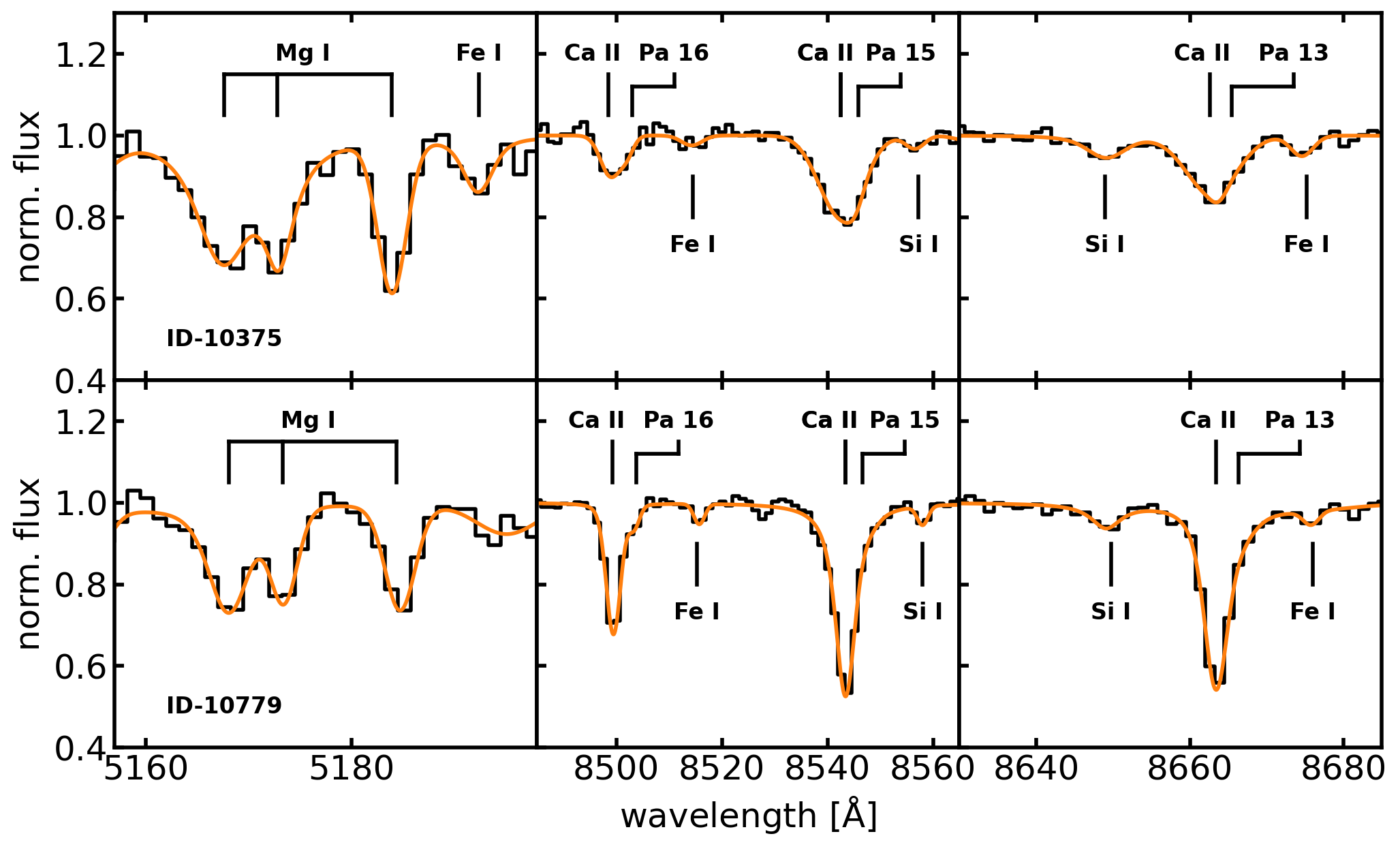}
\caption{The fitted spectrum for star 10375 (top row) and 10779 (bottom row). The extracted and rectified stellar spectrum in the regions of the \ion{Mg}{1} and \ion{Ca}{2} triplet is shown in black and with the best spectral fit in orange.}
\label{fig:spec_fit_MgICaII}
\end{figure}

\begin{figure}[htb]
\plottwo{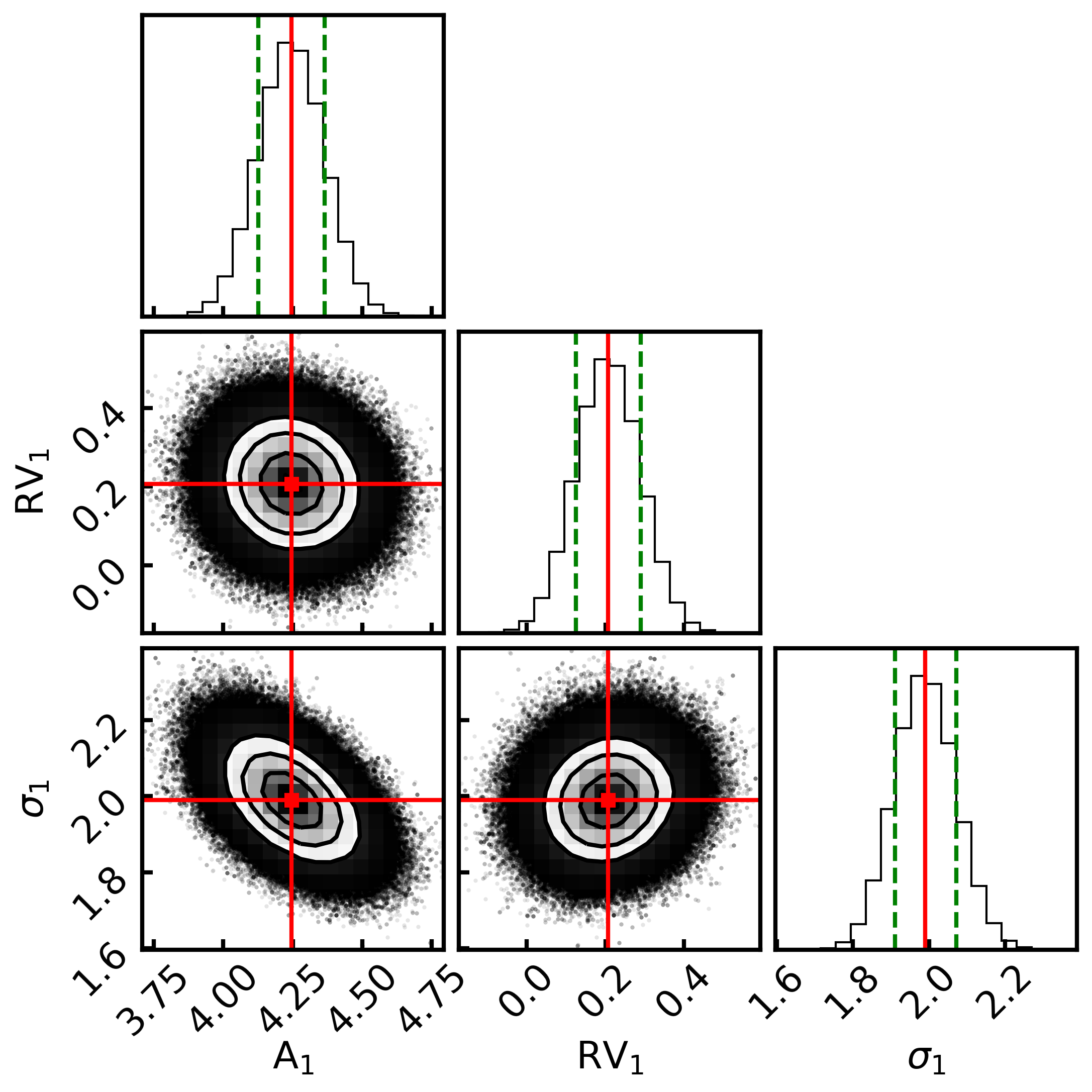}{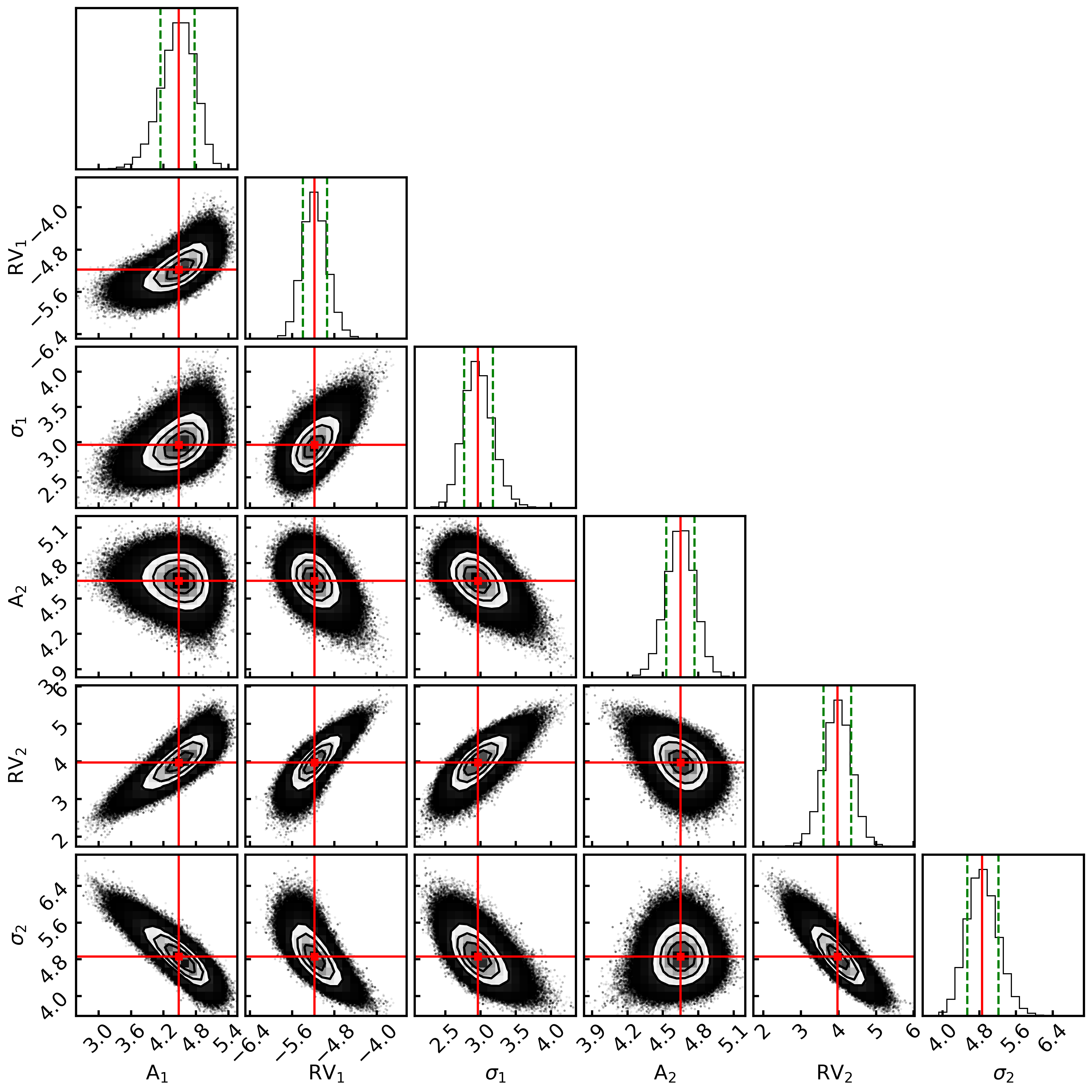}
\caption{The corner plot of the MCMC run of the O-star (left) and B-star (right) RV distribution. The corresponding results ($50^{\rm th}$ percentile) are marked as red lines and additionally in the histograms the $1\sigma$ uncertainties ($84^{\rm th}$ percentile) are marked as dashed green lines.}
\label{fig:corner_obstars}
\end{figure}

\begin{figure}[htb]
\plotone{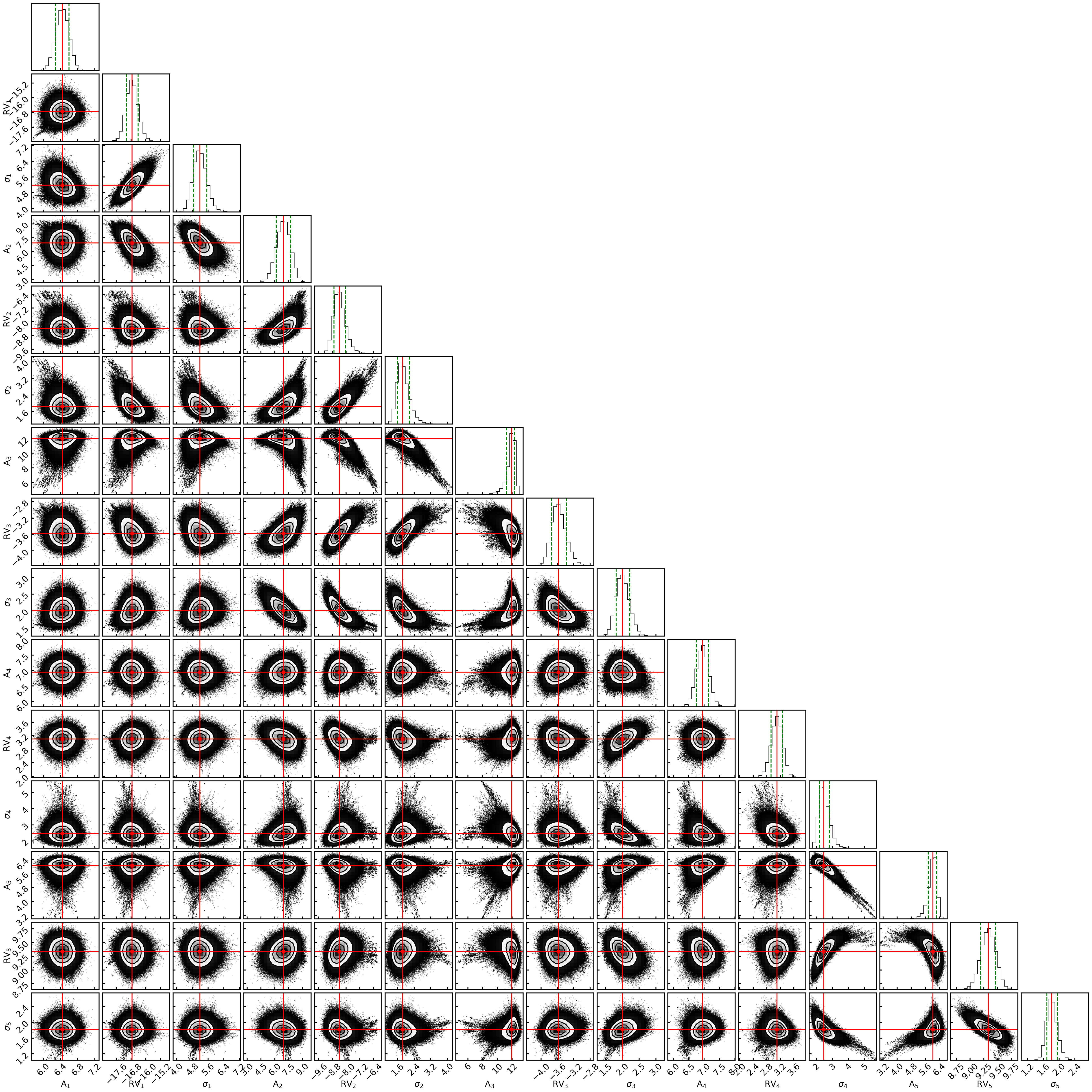}
\caption{Same as Fig.~\ref{fig:corner_obstars} for the PMS stars}
\label{fig:corner_pms}
\end{figure}

\begin{figure}[htb]
\epsscale{0.6}
\plotone{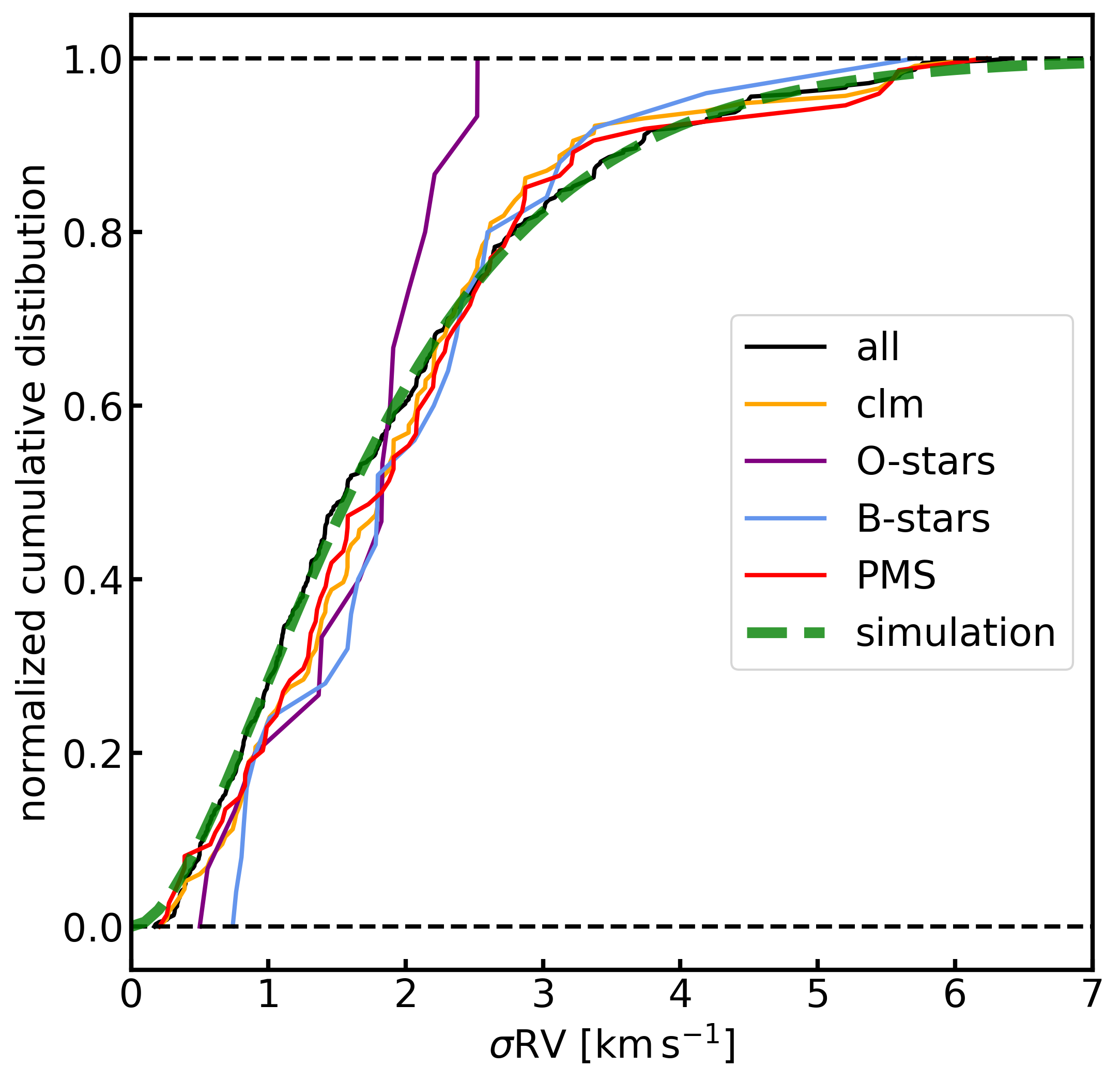}
\caption{The normalized cumulative distribution of the RV uncertainties used to define the statistical error distribution of errors. The dashed green lines represents the model used to sample the RV uncertainties.}
\label{fig:err_dist}
\end{figure}

\section{Tables}

\begin{deluxetable}{rrrllrrrrrrrr}[htb]
	\tablecaption{The runaway star candidates\label{tab:runaways}}
	\tablehead{\multicolumn{1}{c}{ID} &\multicolumn{2}{c}{coordinates} & \multicolumn{1}{c}{\textit{F555W}} & \multicolumn{1}{c}{\textit{F814W}} & \multicolumn{1}{c}{$v_{\rm pec}$} & \multicolumn{1}{c}{$\sigma v_{\rm pec}$} & \multicolumn{1}{c}{RV} & \multicolumn{1}{c}{$\sigma$RV} &\multicolumn{1}{c}{$\mu_{\alpha \ast}$} &\multicolumn{1}{c}{$\sigma_{\alpha \ast}$} & \multicolumn{1}{c}{$\mu_{\delta}$} & \multicolumn{1}{c}{$\sigma_{\delta}$}\\
	\multicolumn{1}{c}{}&\multicolumn{1}{c}{r.a.}&\multicolumn{1}{c}{dec.} & \multicolumn{1}{c}{(mag)} & \multicolumn{1}{c}{(mag)} &\multicolumn{8}{c}{(${\rm km}\,{s}^{-1}$)}}
	\startdata
4644 & $10^\mathrm{h}23^\mathrm{m}58.21^\mathrm{s}$ & $-57^\circ44{}^\prime19.25{}^{\prime\prime}$ & 22.186 & 18.320 & 56.2 & 25.2 & -7.55 & 2.15 & -20.1 & 23.9 & 51.9 & 21.5 \\
5216 & $10^\mathrm{h}23^\mathrm{m}58.93^\mathrm{s}$ & $-57^\circ46{}^\prime00.83{}^{\prime\prime}$ & 19.725 & 17.474 & 76.3 & 7.8 & 46.80 & 1.46 & -60.2 & 7.4 & -0.0 & 7.4 \\
6187 & $10^\mathrm{h}23^\mathrm{m}59.98^\mathrm{s}$ & $-57^\circ46{}^\prime17.86{}^{\prime\prime}$ & 21.635 & 18.565 & 36.2 & 23.3 & 34.30 & 1.08 & -10.7 & 20.6 & -4.1 & 23.2 \\
6708 & $10^\mathrm{h}24^\mathrm{m}00.50^\mathrm{s}$ & $-57^\circ45{}^\prime15.78{}^{\prime\prime}$ & 19.102 & 16.414 & 33.7 & 5.3 & 33.12 & 2.35 & -3.4 & 4.6 & -5.2 & 4.3 \\
7370 & $10^\mathrm{h}24^\mathrm{m}01.13^\mathrm{s}$ & $-57^\circ45{}^\prime21.41{}^{\prime\prime}$ & 20.025 & 16.783 & 40.6 & 10.0 & -27.64 & 2.75 & -29.7 & 9.6 & -1.2 & 8.8 \\
7834 & $10^\mathrm{h}24^\mathrm{m}01.55^\mathrm{s}$ & $-57^\circ44{}^\prime05.67{}^{\prime\prime}$ & --- & $12.094^{\rm a}$ & 59.5 & 0.9 & 6.02 & 0.32 & -57.8 & 0.8 & -13.0 & 0.7 \\
8605 & $10^\mathrm{h}24^\mathrm{m}02.23^\mathrm{s}$ & $-57^\circ45{}^\prime56.26{}^{\prime\prime}$ & 19.353 & 16.783 & 56.6 & 6.3 & -18.09 & 0.58 & -46.6 & 5.8 & 26.6 & 6.2 \\
8707 & $10^\mathrm{h}24^\mathrm{m}02.32^\mathrm{s}$ & $-57^\circ45{}^\prime41.74{}^{\prime\prime}$ & 19.127 & 16.083 & 38.1 & 18.9 & -28.24 & 0.99 & 23.5 & 12.2 & 10.0 & 15.0 \\
10048 & $10^\mathrm{h}24^\mathrm{m}03.59^\mathrm{s}$ & $-57^\circ45{}^\prime27.08{}^{\prime\prime}$ & 16.809 & 14.551 & 32.8 & 4.1 & 3.65 & 1.01 & -15.5 & 3.8 & 28.7 & 3.9 \\
10198 & $10^\mathrm{h}24^\mathrm{m}03.77^\mathrm{s}$ & $-57^\circ44{}^\prime39.79{}^{\prime\prime}$ & 15.414 & 13.328 & 43.3 & 1.5 & 42.99 & 0.50 & 1.3 & 1.2 & -5.3 & 1.2 \\
10779 & $10^\mathrm{h}24^\mathrm{m}04.42^\mathrm{s}$ & $-57^\circ46{}^\prime03.65{}^{\prime\prime}$ & 19.960 & 17.545 & 54.9 & 11.7 & 28.49 & 0.61 & 45.8 & 10.3 & -10.5 & 10.8 \\
10924 & $10^\mathrm{h}24^\mathrm{m}04.57^\mathrm{s}$ & $-57^\circ46{}^\prime41.19{}^{\prime\prime}$ & 15.313 & 13.668 & 50.1 & 1.1 & -31.63 & 0.36 & 36.8 & 1.0 & 12.4 & 1.0 \\
13587 & $10^\mathrm{h}24^\mathrm{m}07.91^\mathrm{s}$ & $-57^\circ45{}^\prime22.69{}^{\prime\prime}$ & 18.756 & 16.080 & 123.2 & 4.2 & 44.31 & 0.39 & -107.2 & 4.2 & 41.6 & 4.0 \\
14161 & $10^\mathrm{h}24^\mathrm{m}08.74^\mathrm{s}$ & $-57^\circ44{}^\prime43.97{}^{\prime\prime}$ & 19.653 & 17.121 & 36.3 & 8.8 & -17.35 & 0.66 & 26.4 & 7.5 & -17.9 & 8.8 \\
14542 & $10^\mathrm{h}24^\mathrm{m}09.22^\mathrm{s}$ & $-57^\circ43{}^\prime57.67{}^{\prime\prime}$ & 18.530 & 16.675 & 546.1 & 5.3 & -6.84 & 2.62 & -289.6 & 3.6 & 462.9 & 4.3 \\
14581 & $10^\mathrm{h}24^\mathrm{m}09.26^\mathrm{s}$ & $-57^\circ45{}^\prime27.56{}^{\prime\prime}$ & 22.037 & 18.882 & 89.2 & 34.6 & -13.56 & 2.47 & 0.3 & 28.2 & 88.1 & 33.1 \\
15613 & $10^\mathrm{h}24^\mathrm{m}10.69^\mathrm{s}$ & $-57^\circ46{}^\prime00.00{}^{\prime\prime}$ & 16.387 & 12.783 & 30.9 & 1.9 & -21.65 & 0.20 & 9.1 & 1.8 & -20.0 & 1.8 \\
15956 & $10^\mathrm{h}24^\mathrm{m}11.22^\mathrm{s}$ & $-57^\circ45{}^\prime45.55{}^{\prime\prime}$ & 21.533 & 18.410 & 57.9 & 29.0 & 2.98 & 5.53 & -38.6 & 27.9 & 43.1 & 21.1 \\
16306 & $10^\mathrm{h}24^\mathrm{m}11.74^\mathrm{s}$ & $-57^\circ45{}^\prime18.94{}^{\prime\prime}$ & 17.390 & 15.089 & 245.8 & 2.3 & 129.08 & 0.83 & 189.1 & 2.0 & -89.5 & 2.0 \\
16549 & $10^\mathrm{h}24^\mathrm{m}12.14^\mathrm{s}$ & $-57^\circ44{}^\prime44.84{}^{\prime\prime}$ & 19.866 & 17.587 & 54.9 & 9.1 & -29.75 & 1.91 & -45.4 & 8.7 & 8.6 & 8.7 \\
17635 & $10^\mathrm{h}24^\mathrm{m}14.15^\mathrm{s}$ & $-57^\circ44{}^\prime20.27{}^{\prime\prime}$ & 18.866 & 17.033 & 75.3 & 5.7 & 74.31 & 1.73 & 10.1 & 5.4 & -6.6 & 5.4 \\
19384 & $10^\mathrm{h}24^\mathrm{m}18.52^\mathrm{s}$ & $-57^\circ46{}^\prime41.56{}^{\prime\prime}$ & 18.607 & 15.887 & 68.0 & 3.5 & 3.96 & 0.69 & -45.4 & 3.4 & 50.5 & 3.2 \\
	\enddata
	\tablecomments{The runaway candidate stars of Wd2. In Column 1 we list the unique object ID from our photometric HST catalog. Columns~2 and 3 are the coordinates and Columns~4 and 5 show the HST \textit{F555W} and \textit{F814W} magnitude, respectively ($^{\rm a}$this star is saturated in the HST observations and the \textit{F814W} magnitude was recovered from the MUSE data cube). Columns~6 and 7 show the peculiar stellar velocity and its uncertainty. In Columns~8 through 13 we list the individual velocity components and their uncertainties. All velocities are shown in ${\rm km}\,{\rm s}^{-1}$ and the corresponding systemic velocities (${\rm RV}_{\rm sys} = 15.9\,{\rm km}\,{\rm s}^{-1}$, $\mu_{\alpha \ast,{\rm sys}} = -101.9\,{\rm km}\,{\rm s}^{-1}$, $\mu_{\delta,{\rm sys}} = 59.1\,{\rm km}\,{\rm s}^{-1}$) are subtracted.}
\end{deluxetable}


\bibliography{references}{}
\bibliographystyle{aasjournal}



\end{document}